\documentclass[12pt]{article}

\usepackage{amsmath}
\usepackage{graphicx}

\newcommand{\mys}[1]{\section{#1}
	\setcounter{equation}{0}}
	\renewcommand{\theequation}{\arabic{section}.\arabic{equation}}

\newcommand{\myappendix}{
\appendix
\vspace{30pt}
\noindent {\bf \Large Appendix}
\renewcommand{\theequation}{\Alph{section}.\arabic{equation}}
}

\newlength{\dummysp}
\newcommand{\bbar}[1]{{\overline{#1}}}
\newcommand{\half}{{\frac{1}{2}}}

\newcommand{\beq}{\begin{eqnarray}}
\newcommand{\eeq}{\end{eqnarray}}
\newcommand{\nnn}{ \nonumber \\ }
\newcommand{\p}{{\partial}}

\newcommand{\Zbf}{{{\bf Z}}}

\newcommand{\e}{{\epsilon}}
\newcommand{\s}{{\sigma}}
\newcommand{\vev}[1]{{\langle #1 \rangle}}

\newcommand{\ord}[1]{{{\cal O}(#1)}}
\newcommand{\gappeq}{\mathrel{\rlap {\raise.5ex\hbox{$>$}}
{\lower.5ex\hbox{$\sim$}}}}
\newcommand{\lappeq}{\mathrel{\rlap{\raise.5ex\hbox{$<$}}
{\lower.5ex\hbox{$\sim$}}}}
\newcommand{\myref}[1]{(\ref{#1})}

\newcommand{\ben}{\begin{enumerate}}
\newcommand{\een}{\end{enumerate}}

\newcommand{\fourth}{\frac{1}{4}}
\newcommand{\sbar}{{\bar \s}}
\newcommand{\phib}{{\bar \phi}}

\newcommand{\psib}{{\bar \psi}}

\newcommand{\bit}{\begin{itemize}}
\newcommand{\eit}{\end{itemize}}

\newcommand{\mbf}{{\bf m}}

\newcommand{\Ncal}{{\cal N}}

\newcommand{\zb}{{\bbar{z}}}

\newcommand{\be}{{\bar \e}}
\newcommand{\eb}{{\bar \e}}
\newcommand{\Fb}{{\bar F}}

\newcommand{\Qb}{{\bar Q}}

\newcommand{\Wbar}{{\bbar{W}}}

\def\[{\left [}
\def\]{\right ]}
\def\({\left (}
\def\){\right )}

\begin{document}

\begin{titlepage}

\renewcommand{\thefootnote}{\fnsymbol{footnote}}

\hfill July 14, 2005

\hfill hep-lat/0507016

\vspace{0.45in}

\begin{center}
{\bf \Large R-symmetry in the Q-exact (2,2) \\ \vskip 10pt
2d lattice Wess-Zumino model}
\end{center}

\vspace{0.15in}

\begin{center}
{\bf \large Joel Giedt\footnote{{\tt giedt@physics.utoronto.ca}}}
\end{center}

\vspace{0.15in}

\begin{center}
{\it University of Toronto, Department of Physics \\
60 St. George St., Toronto ON M5S 1A7 Canada}
\end{center}

\vspace{0.15in}

\begin{abstract}
In this article we explore the R-symmetry
of the (2,2) 2d Wess-Zumino model.
We study whether or not this symmetry is approximately
realized in the Q-exact lattice version
of this theory.  Our study is nonperturbative:
it relies on Monte Carlo simulations
with dynamical fermions.
Irrelevant operators in the lattice
action explicitly break the R-symmetry.
In spite of this, it is found to be a symmetry of
the effective potential.  We find nonperturbative 
evidence that the nonrenormalization theorem of
the continuum theory is recovered in the
continuum limit; e.g., there is no
additive mass renormalization.  In our simulations
we find that Fourier acceleration of the
hybrid Monte Carlo algorithm allows us
to avoid difficulties with critical slowing-down.
\end{abstract}

\end{titlepage}

\renewcommand{\thefootnote}{\arabic{footnote}}
\setcounter{footnote}{0}

\mys{Motivation}
The continuum (2,2) 2d Wess-Zumino (2dWZ) model
(obtained from a dimensional reduction
of the 4d Wess-Zumino model \cite{Wess:1973kz})
is supposed to provide a Landau-Ginzburg
description of the minimal discrete series
of $\Ncal=2$ superconformal field theories~\cite{Boucher:1986bh}.
In the present article, we examine an important
aspect of the simplest of these models---the 
one with a cubic superpotential---in
the context of class of lattice
actions that have an exact lattice 
supersymmetry.
These lattice actions were first formulated 
in \cite{Elitzur:1983nj,Sakai:1983dg} using 
{\it Nicolai map} \cite{Nicolai:1979nr} methods, 
relying on earlier Hamiltonian \cite{Elitzur:1982vh}
and continuum \cite{Cecotti:1982ad} studies
that also utilized the Nicolai map.
Detailed studies of the spacetime lattice system
were performed in \cite{Beccaria:1998vi} by stochastic
quantization methods and in \cite{Catterall:2001wx}
by the Monte Carlo simulation approach.

Once auxiliary fields are introduced, the
lattice action takes a Q-exact form: $S=QX$,
as was emphasized in topological
interpretation of \cite{Catterall:2003xx}
and the lattice superfield approach of \cite{Giedt:2004qs}.  
Here $Q$ is a lattice supercharge with
derivatives realized through discrete difference
operators; with respect to a
discrete approximation of the continuum
theory superalgebra, $Q^2=0$ is a nilpotent
subalgebra.  Because $S$ is Q-exact, the action is
trivially invariant with respect to this
lattice supersymmetry:  $QS=Q^2X=0$.

In the massive continuum theory, 
as will be shown below,
there is an exact $Z_2(R)$
symmetry.  It is an {\it R-symmetry,}
meaning it does not commute with the supercharges.
This symmetry is spontaneously broken at infinite
volume.  In the massless case, i.e., in the critical
domain, the classical R-symmetry
is enlarged to $U(1)_R$.  It cannot be spontaneously
broken since it is a continuous symmetry in 2d~\cite{MWC}.
If the lattice theory has the correct continuum
limit, it should reproduce these features.  On
the other hand, these R-symmetries are only
approximate in the Q-exact lattice action; the
symmetry is explicitly broken by the Wilson
mass term that is used to lift doublers.\footnote{This
is directly related to the breaking of
the so-called $U(1)_V$ symmetry, that was
pointed out in \cite{Giedt:2004qs}.}

It has been shown in \cite{Giedt:2004qs} that 
the continuum limit
of the lattice perturbation series is identical
to that of the continuum theory,
due to cancellations that follow from $Q^2=0$.
Thus, the Q-exact spacetime lattice has
behavior that is similar to what was
found on the $Q,Q^\dagger$-preserving
spatial lattice in \cite{Elitzur:1983nj}.
However, it was also shown in \cite{Giedt:2004qs}
that the most general continuum
effective action that is consistent
with the symmetries of the bare lattice action
is not the (2,2) 2d Wess-Zumino model.
This raises the question of whether or not the good behavior
of perturbation theory persists at a nonperturbative level.
The Monte Carlo simulation results of 
Catterall et al.~give hope that
the desired continuum limit
is obtained beyond perturbation 
theory~\cite{Catterall:2001wx}.  
If so, this would be one of the few
examples of a supersymmetric field theory that
can be latticized and studied nonperturbatively
by Monte Carlo simulation
without the need for fine-tuning
of counterterms.

Though not directly related, we pause to
mention that lattice super-Yang-Mills theories exist that do
not require fine-tuning of counterterms \cite{Cohen:2003xe}, 
at least within perturbation theory.
However, in some of those cases, it is known
that the fermion determinant has a complex
phase that depends on the boson configuration \cite{Giedt:2003ve}.  
The complex phase
poses a severe difficulty for
Monte Carlo simulations of those systems~\cite{Giedt:2004tn}.

If we can show that
features of the continuum theory associated with
the R-symmetry are recovered in the continuum
limit, it provides further 
evidence that the correct theory is
obtained.  The symmetry that we study
persists in the infrared effective theory
in a strongly coupled regime.  Thus,
we are testing aspects
of the lattice theory that lie beyond
the reach of perturbation theory.

We now summarize the rest of this article:

\bit
\item
In Section \ref{contact}, we briefly describe the continuum
theory that the lattice model is supposed to define.
We discuss both the general (2,2) 2d Wess-Zumino model,
and the specific case that we select for further study:
the model with a cubic superpotential.
In that case, we show that there is a classical
$Z_2(R)$ symmetry.  We explain that it
must be spontaneously broken, according to
a well-known theorem for 2d field theories
with a stable potential.
\item
In Section \ref{s:lat}, the lattice action is described.
Symmetry aspects of the action are emphasized.  In
particular, we show that when doublers are lifted
by a supersymmetric version of the Wilson mass
term, the $Z_2(R)$ is explicitly broken.  However,
it is an approximate classical symmetry for the 
long wavelength modes.
\item
In Section \ref{s:z2r}, 
observables are defined
that are used in our study of the critical
domain.  Basic data regarding their behavior
is briefly described.  Degenerate minima 
of the effective potential lead to 
tunneling in the simulations at
finite volume.  We describe how observables 
are chosen that are insensitive to this
finite volume effect.  
Next the $Z_2(R)$ symmetry is examined in our
simulations.  We find that the effective potential
exhibits this
symmetry to a very good approximation.
We show that as the volume is increased,
spontaneous symmetry-breaking of $Z_2(R)$ is
clearly evident.  
Finally, we demonstrate
that the $U(1)_R$ symmetry of the effective
potential is recovered in the critical domain.
We find evidence that the nonrenormalization
theorems of the continuum theory hold
to a good approximation in the limit
of small lattice spacing.
\item
In Section \ref{s:ifd}, we conclude with an
interpretation of our simulation results.  We
also outline future directions for research
in this lattice system.
\item
In Appendix \ref{sec:sim}, the simulation methods
are outlined.  We describe the extent to which
we were able to overcome critical slowing-down
using Fourier acceleration methods, and some
specifics about the algorithm in this regard.
\eit

\mys{The target theory}
\label{contact}

We begin our discussion by considering the
continuum theory.  Our focus will be on
the simplest (2,2) 2d WZ model:  the one 
with a cubic superpotential interaction.
It is a supersymmetric cousin to an ordinary 
2d complex scalar field $\phi$ model.
Two 2d Majorana fermions $(\psi_-,\psib_-)$ 
and $(\psi_+,\psib_+)$ fit into the
same (2,2) supersymmetry multiplet as $\phi$.

\subsection{The action}

The Euclidean action that we study is
\beq
S &=& \int d^2z \; \[ - 4 \phib \p_z \p_\zb \phi 
-2i \psib_- \p_\zb \psi_- + 2i \psi_+ \p_z \psib_+ 
- \Fb F \right. \nnn
&& \left. + W'(\phi) F + \Wbar'(\phib) \Fb
- W''(\phi) \psi_+ \psi_- - \Wbar''(\phib) \psib_- \psib_+ \]
\label{cact}
\eeq
In these expressions, $z=x_1+ix_2=(\zb)^*$,
$\p_z = (\p_1 - i \p_2)/2 = (\p_\zb)^*$, $d^2z = dz \, d\zb$,
$F$ and $\phi$ are complex scalar fields,
$W'(\phi)= \p W/\p \phi$, etc., where $W(\phi)$ is
holomorphic in $\phi$ and is the {\it superpotential.}
$\psi_\pm$ and $\psib_\pm$ are
independent Grassmann fields:  the partition
function is $Z = \int [d^2\phi \, d^2F \, d^2 \psi_- \, 
d^2 \psi_+] \; \exp(-S)$, where $d^2 \psi_- = d \psi_-
\, d \psib_-$, etc.

Although the quadratic term of the auxiliary
field $F$ has the ``wrong'' sign, it can be
formally integrated by analytic continuation.
This allows for the elimination of $F$,
leading to the action
\beq
S &=& \int d^2z \; \[ - 4 \phib \p_z \p_\zb \phi
-2i \psib_- \p_\zb \psi_- + 2i \psi_+ \p_z \psib_+ 
\right. \nnn
&& \left. + |W'(\phi)|^2 
- W''(\phi) \psi_+ \psi_- - \Wbar''(\phib) \psib_- \psib_+ \]
\label{poiut}
\eeq
The continuum partition function is thus
$Z = \int [d^2 \phi \, d^2 \psi_- \, d^2 \psi_+ ] \exp(-S)$,
now with action~\myref{poiut}.

Of course $Z$ is at this point only formally defined.
Meaning can be assigned to it
through perturbation theory about $\phi=0$, or
some more general set of classical saddlepoints.
In either case, one obtains an asymptotic
series that provides approximate results
valid in a limited regime.  However it is
worth noting that the supersymmetry of the model
leads to nonrenormalization theorems that
allow for a number of exact results to be
obtained; see for example \cite{Witten:1993jg}
and references therein.
It is hoped that the lattice formulation that we
describe below will give a more complete meaning
to the functional integral, outside of perturbation
theory and semiclassical expansion.
In that case, aspects of the theory that
are not protected by nonrenormalization
theorems can be studied in a strongly-coupled
or deep infrared regime.

We specialize to the superpotential
\beq
W(\phi) = \frac{m}{2} \phi^2 + \frac{g}{3!} \phi^3
\label{cspot}
\eeq
Both $m$ and $g$ have mass dimension 1.  Perturbation
theory is defined by $g/m \ll 1$.  It is
useful to make the field redefinition
\beq
\phi = -\frac{m}{g} + \s
\label{sfld}
\eeq
In terms of $\s$, the superpotential becomes
\beq
W(\s) = - \lambda \s + \frac{g}{3!} \s^3, \quad
\lambda = \frac{m^2}{2g}
\label{wsig}
\eeq

\subsection{Classical symmetries}

With periodic boundary conditions for all fields,
the action \myref{cact} is invariant under the infinitesmal
supersymmetry transformations (also true
for $\phi \to \s$)
\beq
\delta \phi &=& \e^- \psi_- + \e^+ \psi_+ , \qquad
\delta \psi_+ = -2i \be^+ \p_\zb \phi + \e^- F \nnn
\delta \psi_- &=&  2i \be^- \p_z \phi - \e^+ F , \qquad
\delta F = 2i \be^- \p_z \psi_+ + 2i \be^+ \p_\zb \psi_- \nnn
\delta \phib &=& - \eb^- \psib_- - \eb^+ \psib_+ , \qquad
\delta \psib_+ = 2i \e^+ \p_\zb \phib + \eb^- \Fb \nnn
\delta \psib_- &=& -2i \e^- \p_z \phib - \eb^+ \Fb , \qquad
\delta \Fb = 2i \e^- \p_z \psib_+ + 2i \e^+ \p_\zb \psib_-
\label{csutr}
\eeq
It can be seen that $\e^+,\be^+$ are associated
with $\p_\zb$ whereas $\e^-,\be^-$ are associated
with $\p_z$.  We can associate supercharge operators
with the transformations \myref{csutr} through
\beq
\delta \equiv \e^- Q_- + \e^+ Q_+ - \be^- \Qb_-
- \be^+ \Qb_+
\label{qdf}
\eeq
From this definition it is straightforward to work
out the action of $Q_\pm,\Qb_\pm$ on the fields.
Two subalgebras emerge:
\beq
\{ Q_- , \Qb_- \} = -2i \p_z, \quad 
\{ Q_+ , \Qb_+ \} = 2i \p_\zb
\label{22alg}
\eeq
All other anticommutators vanish.
Note that from the relation $t_M= -it_E$
between Minkowski and Euclidean time,
\beq
z=x + it_E = x-t_M, \quad
\zb=x - it_E = x+t_M
\eeq
Thus the first subalgebra in \myref{22alg}
closes on the generator of left-moving translation
and the second subalgebra closes on the
generator of right-moving translation.
Respectively, there is a left-moving (2,0) algebra
and a right-moving (0,2) algebra.  The ``2'' denotes
two generators, say $Q_-$ and $\Qb_-$.
Taken together, the system has (2,2) supersymmetry.

The superpotential \myref{wsig} has a $U(1)_R$
symmetry when $\lambda=0$, $g \not= 0$ (note that
$\phi=\s$ when $\lambda=0$):
\beq
\s \to e^{2 i \alpha /3} \s, \quad 
\psi_\pm \to e^{-i \alpha/3} \psi_\pm, \quad
F \to e^{-4i \alpha/3} F
\label{rtran}
\eeq
and $\sbar,\psib_\pm,\Fb$ transforming as
conjugates.  Note that $\s$ has R-charge
$2/3$.  Thus $W \to \exp(2i\alpha) W$,
which implies that the superpotential
(still with $\lambda=0$) has R-charge~2;
this is just the well-known condition to have a $U(1)_R$
symmetry.  That this is an R-symmetry follows from the
fact that the component fields of the
same supermultiplet have different R-charges,
as can be seen from \myref{rtran}.

There is also an axial $U(1)_A$ symmetry that we will
make use of:
\beq
\psi_\pm \to e^{\pm i \omega} \psi_\pm, \quad
\psib_\pm \to e^{\mp i \omega} \psi_\pm
\label{u1a}
\eeq
with all other fields neutral.  Note that this
contains fermion parity $Z_2(F)$ as a subgroup.

We now show that the classical theory with $\lambda \not=0$,
$g \not= 0$ has an exact $Z_2(R)$ symmetry.
The scalar potential is
\beq
V = |W'(\s)|^2 = \left| \lambda - \frac{g}{2} \s^2 \right|^2
\label{spt}
\eeq
This potential has minima (recall $\lambda=m^2/2g$)
\beq
\s_\pm = \pm \sqrt{2\lambda/g} =\pm m/g
\label{clmin}
\eeq
Clearly \myref{spt} has a $Z_2$ symmetry $\s \to - \s$
that relates the two classical vacua.
The $Z_2$ is a subgroup of the $U(1)_R \times U(1)_A$ that survives
when $\lambda \not= 0$ in \myref{wsig}, as will
be discussed below.
Since $V=|W'(\s)|^2 \geq 0$, these are absolute minima
with $V(\s_\pm)=0$.  The saddle point
that separates these minima is at the
origin, with
\beq
W''(0)=0, \quad V(0) = |\lambda|^2=\frac{|m|^4}{4 |g|^2}
\eeq
It can be seen that as $|g|$ is decreased with
$m$ held fixed, the minima
separate and the energy density of
the barrier between them increases
in height.

The effect of $\lambda \not= 0$ is best understood by
treating it as a background field with
nonvanishing R-charge.  For the
superpotential to have the requisite
R-charge of 2, $\lambda$ must have R-charge
of $4/3$:
\beq
\lambda \to e^{4i \alpha/3} \lambda
\eeq
Now notice that $\lambda \not= 0$ is left invariant
iff
\beq
\alpha = \frac{3 \pi n}{2}, \quad n \in \Zbf
\eeq
Thus nonzero $\lambda$ leaves intact the $Z_4(R)$
subgroup generated by $\alpha=3 \pi/2$
in Eq.~\myref{rtran}.  At the
level of component fields, this tranformation 
is $(\omega \equiv e^{i \pi/4})$:
\beq
&& \s \to \Omega(\omega) \s = \omega^2 \s = -\s,
\quad \psi_\pm \to \Omega(\omega) \psi_\pm =
\omega^{-1} \psi_\pm = -i \psi_\pm \nnn
&& F \to \Omega(\omega) F = F
\label{z4t}
\eeq
and $\sbar,\psib_\pm,\Fb$ transforming as
conjugates.  Here, $\Omega$ is the homomorphism
that determines how the $Z_4(R)$ acts on the fields;
i.e., their 4-ality.  It is easily checked that
this is a symmetry of the full
action with superpotential \myref{wsig}.

To obtain the promised $Z_2(R)$, we combine
the transformation \myref{z4t} with a $Z_4(A)$
subgroup of the $U(1)_A$ that appeared in
in \myref{u1a}.  We describe it by the action
of a generator $\gamma_3$:
\beq
\gamma_3: \quad \psi_+ \to i \psi_+,
\quad \psi_- \to -i \psi_-, 
\quad \psib_+ \to -i \psib_+, 
\quad \psib_- \to i \psib_-
\eeq
The $Z_2(R)$ is generated by $\gamma_3 \Omega(\omega)$:
\beq
Z_2(R): \quad \s \to - \s, \quad \sbar \to - \sbar, 
\quad \psi_\pm \to \pm \psi_\pm,
\quad \psib_\pm \to \pm \psib_\pm
\label{z2r}
\eeq
Thus for $\lambda \not= 0, g \not=0$, or equivalently
$m \not=0, g \not= 0$,
the surviving global symmetry is $U(1)_A \times
Z_4(R) \equiv U(1)_A \times Z_2(R)$.

At infinite volume,
the classical $Z_2(R)$ symmetry must be spontaneously
broken.  This is because
the scalar potential is positive semi-definite:
$V=|W'(\s)|^2$.  In this case, the stability of
the scalar potential implies in 2d that there are
no nontrivial finite action solutions:  only
the constant absolute minima $\s_\pm$ 
of \myref{clmin} have finite 
action.\footnote{See for example Section 41.4 of \cite{ZJ}.}
Thus instantons do not exist, tunneling does
not occur, and the symmetry is spontaneously
broken.

\subsection{Renormalization}

The only renormalization of the superpotential
$W$ is that due to wavefunction 
renormalization \cite{Wess:1973kz}, with
all component fields in the same
supermultiplet rescaled identically.\footnote{Of course,
Wess and Zumino proved the nonrenormalization
theorem in 4d; but the crucial
cancellations, due to the 4 conserved 
supercharges, continue to hold in 2d.}
Thus if the bare fields and renormalized fields
are related by $\phi= \sqrt{Z} \phi_r$, $\psi_{\pm} = \sqrt{Z}
\psi_{\pm,r}$, etc, then the renormalized
superpotential is completely described by the
identification $W(m,g|\phi) \equiv W(m_r,g_r|\phi_r)$ with
\beq
m_r = Z m, \quad
\quad g_r = Z^{3/2} g
\label{wfren}
\eeq
Note that this implies $\lambda_r = \sqrt{Z} \lambda$.
Counterterms $\delta m$ and $\delta g$ 
for the couplings are not required.  I.e.,
there is no additive renormalization of $m$ or $g$;
if the bare mass $m=0$, then the
theory remains massless under renormalization.
This can also be seen as a consequence of
the $U(1)_R$ symmetry \myref{rtran} that
occurs if $m=0$.  Since a continuous symmetry
cannot be broken in 2d \cite{MWC},
we know that a mass
term cannot be radiatively generated, since
that would break the $U(1)_R$ symmetry.
Thus in the continuum theory the critical
domain is the neighborhood of $m=0$ for
any $g$.  We will find that this is likewise
true in the lattice theory.

\mys{The lattice action}
\label{s:lat}

We now describe the Q-exact lattice action that
is the subject of our Monte Carlo simulation studies.  
The general form of this action
was first constructed in \cite{Elitzur:1983nj,Sakai:1983dg} using 
a Nicolai map \cite{Nicolai:1979nr} approach.
In \cite{Catterall:2001wx}, Monte Carlo simulations were performed.
These demonstrated that Ward identities associated
with the (2,2) supersymmetry of the continuum theory
were satisfied to such a good approximation
that no statistically significant violation could
be observed, even with a large sample size.
In \cite{Beccaria:1998vi} and \cite{Catterall:2001wx},
two different methods demonstrated boson-fermion
spectrum degeneracy for the lightest states, 
a necessary condition for supersymmetry.
Furthermore, the numerical values for the light spectrum
agreed with those obtained on the
$Q,Q^\dagger$-preserving spatial lattice in \cite{Elitzur:1983nj}.
In \cite{Giedt:2004qs} the lattice
action was derived using a lattice superfield approach.
It was shown how to introduce a Wilson mass term
in the superpotential
to lift doublers, while keeping the one exact
supersymmetry.  The same Wilson mass term also
appeared in the Nicolai map construction of
\cite{Elitzur:1983nj,Sakai:1983dg}.  In \cite{Giedt:2004qs}
it was shown that in the $a \to 0$ limit 
the lattice perturbation
series goes over to that of the continuum
without the need for any counterterms,
to all orders in the coupling $g$.

With auxiliary fields eliminated, 
and implicit summation over repeated
lattice site indices $m,n$, the action is:\footnote{In
all of our expressions we work in the natural
lattice units $a=1$, unless otherwise noted.}
\beq
S &=& 4 [\Delta_z^S \phib + \frac{i}{2} W'(\phi) ]_m
[\Delta_\zb^S \phi - \frac{i}{2} \bbar{W}'(\phib) ]_m
- 2i \psib_{-,m} \Delta_{\zb}^S \psi_{-,m} \nnn
&& + 2i \psi_{+,m} \Delta_z^S \psib_{+,m}
- \psi_{+,m} W''(\phi)_{m n} \psi_{-,n}
- \psib_{-,m} \Wbar''(\phib)_{m n} \psib_{+,n}
\label{lact}
\eeq
For convenience, we define the superpotential
to include a sum over lattice sites:
\beq
W = \sum_m \( - \frac{r}{4} \phi_m \Delta^2 \phi_m
+ \sum_{n>0} \frac{g_n}{n!} \phi_m^n \)
\label{gspt}
\eeq
The purpose of the Wilson mass term $\phi \Delta^2 \phi$, 
which includes nearest neighbor interactions, is to lift doublers in
the spectrum in a supersymmetric way.  From \myref{gspt}
one obtains
\beq
W'(\phi)_m = \frac{\p W}{\p \phi_m}, \quad
W''(\phi)_{m n} = \frac{\p^2 W}{\p \phi_m \p \phi_n}
\eeq
In expressions \myref{lact} and \myref{gspt}, 
the following finite difference
operators are used:
\beq
&& \Delta_\mu^S = \half \( \Delta^+_\mu + \Delta^-_\mu \), 
\qquad \Delta^2 = \sum_{\mu=1,2} \Delta^+_\mu \Delta^-_\mu 
\nnn 
&& 
\Delta^S_z = \half \( \Delta^S_1 - i \Delta^S_2 \), \qquad
\Delta^S_\zb = \half \( \Delta^S_1 + i \Delta^S_2 \)
\label{jsrk}
\eeq
where $\Delta_\mu^+$ and $\Delta_\mu^-$ are the
usual forward and backward difference operators
respectively.
From now on we drop the superscript $S$, leaving it implied,
except on the operator
\beq
\Delta^{S2} = \sum_\mu \Delta^S_\mu \Delta^S_\mu =
4 \Delta^S_z \Delta^S_\zb
\eeq
which we want to distinguish from $\Delta^2$.
We will often suppress site indices as well.

We pause to note that \myref{lact} may be obtained
in terms of the one exact lattice supersymmetry 
$Q$ acting on an expression of
component fields.  This $Q$ is the linear combination
$Q = Q_- + \Qb_+$ of continuum 
supersymmetries \myref{csutr}-\myref{qdf},
with $\p_z \to \Delta_z$ and $\p_\zb \to \Delta_\zb$.
Note that
\beq
Q^2 = \half \{Q_- + \Qb_+ , Q_- + \Qb_+ \} = 0
\label{qan}
\eeq
so that $Q$ is indeed nilpotent.
For reference, the action of 
$Q$ on the component fields is
\beq
&& Q \phi = \psi_-, \quad
Q \psi_+ = F + 2i \Delta_\zb \phi, \quad
Q \psi_- = 0 , \quad
Q F = -2i \Delta_\zb \psi_-, \nnn
&& Q \phib = \psib_+ , \quad
Q \psib_+ = 0 , \quad
Q \psib_- = \Fb - 2i \Delta_z \phib , \quad
Q \Fb = 2i \Delta_z \psib_+
\eeq
from which one can explicitly verify the
property \myref{qan}.
For a general holomorphic function $W'_m(\phi)$ 
we find that
\beq
S &=&  Q \( -F \psib_{-} 
- 2i \psi_{+} \Delta_{z} \phib
+ W'(\phi) \psi_{+} + \bbar{W}'(\phib) 
\psib_{-} \) \nnn 
&=& -2i \psib_- \Delta_\zb \psi_- + 2i \psi_+ \Delta_z \psib_+
- 4 \phib \Delta_z \Delta_\zb \phi - \Fb F 
- \psib_- \bbar{W}'' \psib_+ \nnn
&& \quad - \psi_+ W'' \psi_-
+ W'(\phi) (F + 2i \Delta _\zb \phi)
+ \bbar{W}'(\phib) (\Fb - 2i \Delta_z \phib)
\label{iio}
\eeq
Upon elimination of the auxiliary field $F$
through its equation of motion,
\myref{iio} becomes just the action \myref{lact}.  

Because \myref{iio} is
$Q$-exact, i.e., $S = Q(\cdots)$, we know
that $Q S = 0$.  That is, the exact
supersymmetric invariance of the lattice
action w.r.t.~$Q$ just follows from the
nilpotency of $Q$ and the $Q$-exactness
of the lattice action.  This obviously
provides a mechanism by which a host of
supersymmetric lattice systems can be constructed.
In~\cite{Giedt:2004qs} it was shown that $\myref{lact}$ is far
from the most general action with the symmetries
of $\myref{lact}$.  Hence the importance of
studying the renormalization of the lattice theory.
In particular, we would like to examine
whether or not relations like \myref{wfren} hold.
In our simulation results below,
we will find evidence that \myref{wfren}
is approximately true, in a regime
where $a \ll 1$.

It is a simple matter to specialize to
the case of the cubic superpotential.  Then
\beq
W(\phi) =  \sum_m \( - \frac{r}{4} \phi_m \Delta^2 \phi_m
+ \frac{m}{2} \phi_m^2 + \frac{g}{3!} \phi^3_m \)
\eeq
The action can be written as $S=S_0 + S_{int}$,
where $S_0$ is quadratic in fields:
\beq
S_0 &=& \phib \[ - \Delta^{S2} + 
(\bar m - \frac{\bar r}{2} \Delta^2 )
(m - \frac{r}{2} \Delta^2 ) \] \phi
-2i \psib_{-} \Delta_{\zb} \psi_{-} \nnn
&& + 2i \psi_{+} \Delta_z \psib_{+}
- \psi_{+} (m-\frac{r}{2} \Delta^2) \psi_{-}
- \psib_{-} (\bar m - \frac{\bar r}{2} \Delta^2) \psib_{+}
\quad
\eeq
At $r=0$ the bosons and fermions both have
doublers.  The Wilson mass terms at $r \not= 0$
lifts these modes.  Note that the inverse
free propagator for the boson is the square
of that for the fermions.  This is a consequence
of the exact lattice supersymmetry, and
is a property that is crucial to the
perturbative results that were mentioned above.
The interaction terms are contained in
\beq
S_{int} &=& - g \phi \psi_+ \psi_- 
- \bar g \phib \psib_- \psib_+
+ \half \bar g m \phib^2 \phi + \half g \bar m \phi^2 \phib 
+ \fourth |g|^2 \phi^2 \phib^2 \nnn &&
- \fourth r \bar g \phib^2 \Delta^2 \phi
- \fourth \bar r g \phi^2 \Delta^2 \phib
+ i g \phi^2 \Delta_\zb \phi - i \bar g \phib^2 \Delta_z \phib
\eeq
Note that irrelevant operators appear in the second
line.  These are a consequence of the exact
lattice supersymmetry.  The $r$-dependent
terms originate from the supersymmetrization
of the Wilson mass term.  The other two
arise from crossterms in the first line of \myref{lact},
such as $W'(\phi) \Delta_\zb \phi$.  The crossterms
only contribute to interactions, due
to the identity $\phi \Delta^S_\mu \phi=0$ on
a periodic lattice.  

The irrelevant operators
coming from crossterms violate the $Z_4$ rotation
symmetry of the lattice, as well as reflection
positivity.  As is well-known,
$Z_4$ rotation symmetry is sufficient to
imply Euclidean rotation invariance, $SO(2)$,
of the continuum limit.  Thus the violation
of $Z_4$ rotation symmetry is an important matter,
since it could conceivably destroy Euclidean
invariance of the continuum limit.  It can be shown
\cite{Giedt:2004qs} that the $Z_4$ breaking in
this system is an $\ord{a^2 g_{phys}}$ effect,
where $g_{phys}$ is the coupling in physical
units.  It is harmless to the continuum limit in perturbation
theory, as the lattice perturbation series turns
out to be finite.  However, it remains to be shown
that this does not affect the continuum limit
at a nonperturbative level.  We will defer detailed
discussion of this issue to a later publication.
Suffice it to say, we have found in simulations
that Green function data at more than a few
lattice spacings exhibits approximate
$Z_4$ symmetry, and that this approximation 
improves as one approaches the continuum limit.
As to the lack of reflection positivity,
this too is an $\ord{a^2 g_{phys}}$ effect.  Some problems
were detected in propagators at $g_{phys}=\ord{1}$ in
\cite{Catterall:2001wx}; however, these difficulties
apparently went away for sufficiently small 
lattice spacing $a$.  In the studies that
we perform here, we work in a regime 
where $g = g_{phys} a \ll 1$, where the
violation of reflection positivity by irrelevant
operators is believed to be harmless.

Next, consider the field redefinition \myref{sfld}
in the lattice theory.  One finds
\beq
W(\s) =  \sum_m \( - \frac{r}{4} \s_m \Delta^2 \s_m
- \lambda \s_m + \frac{g}{3!} \s^3_m \)
\eeq
and the lattice action is just \myref{lact} with
$\phi \to \s$.
If $r=0$, then $W'(\s)$ is even and $W''(\s)$
is odd under the $Z_2(R)$ transformation \myref{z2r}.
Taking into account the transformation of the
fermions, the action is by inspection $Z_2(R)$ invariant.
However, the Wilson mass term violates the $Z_2(R)$ symmetry
\myref{z2r} since for $r \not= 0$, $W'(\s)$ and $W''(\s)$
no longer have definite parity with respect
to $Z_2(R)$.  On the other hand, 
the symmetry breaking arises entirely from
irrelevant higher-derivative terms.  Given
that the lattice perturbation series is finite,
the $Z_2(R)$ is restored in the continuum
limit in perturbation theory.  What is left
to check is that this occurs nonperturbatively.
We will find rather strong evidence that this
is true.

Finally, we note that in all of our simulations
we set $r=1$ in the Wilson mass term.

\mys{Simulation results}
\label{s:z2r}

\subsection{Location of the critical domain}

First we introduce observables that allow
us to locate the critical domain.\footnote{Of
course in our analysis we work at finite
volume, so all so-called critical behavior
is actually pseudocritical.  We will not
belabor this point in our terminology
by appending ``pseudo-'' to every instance
of ``critical.''}
For a generic choice of the bare
parameters $m$ and $g$, the effective potential
has two minima, in correspondence with
the classical minima $\s_\pm$ of \myref{clmin}.
We will refer to these two minima of the
effective potential as the $\s_\pm$ vacua,
although their location will be somewhat
shifted from the classical values.
The definitions
that we choose are intended to address the following
issue related to the existence of two vacua.
At finite volume tunneling between the
two vacua $\s_\pm$ will occur.
This is important, because the tunneling events are very
large fluctuations, of order $m/g$.  We do not
want to include such fluctuations in the measurement
of correlations,
since they do not correspond to fluctuations about
one of the $Z_2(R)$ symmetry breaking vacua in
the thermodynamic limit.
By $\s_\pm$ domains we will mean connected regions where
$\s$ is closer to one minimum of the effective
potential than the other.
Several $\s_\pm$ domains may establish themselves on
the $N \times N$ lattice, particularly for large $N$.  In that
case what we really want to measure is the fluctuation
away from the average value within a domain.

Under the assumption of approximate 
$Z_2(R)$ symmetry of the effective potential
at finite volume,
we expect that the following occurs.  Take $m,g$ real
so that the classical minima \myref{clmin} are 
real.  Denote the real and imaginary parts of $\s_m$
as $\s_{R,m}$ and $\s_{I,m}$ respectively.
Suppose we have a ensemble of configurations $\Gamma$.
We partition this ensemble as $\Gamma = \Gamma_+ + \Gamma_- + \Gamma_0$
where: $\s_{R,m} > 0$ on the subensemble $\Gamma_+$,
$\s_{R,m} < 0$ on the subensemble $\Gamma_-$,
and $\s_{R,m}=0$ on the subensemble $\Gamma_0$.
Note that $\Gamma_0$ is a set of measure zero
in any expectation value over $\Gamma$, since
$\s_{R,m}$ is a continuous variable.
Let $\vev{\s_{R}}_+$
be the average of $\s_{R,m}$ over the subensemble $\Gamma_+$
and $\vev{\s_{R}}_-$ be the average
for the subensemble $\Gamma_-$.  The
order parameter $\vev{|\s_R|}$
will denote the average of $|\s_{R,m}|$
over the full ensemble $\Gamma$.  Then we expect
\beq
\vev{\s_{R}}_+ = -\vev{\s_{R}}_- = \vev{|\s_{R}|} \equiv v
\label{vevhyp}
\eeq
to a very good approximation.  The veracity of
this conjecture is borne out in our simulations.
As an example, in Table~\ref{sevev} we show
results for the quantities of \myref{vevhyp} for
$(m,g)=(0.10,0.03)$ and three different
sizes of $N \times N$ lattice.  Similar behavior
was observed at other values of $(m,g,N)$.
This is already good evidence that $Z_2(R)$
is a symmetry of the effective potential,
to within the 1-2\% statistical
uncertainties of the measurements.

\begin{table}
\begin{center}
\begin{tabular}{cccc}
$N$ & $\vev{\s_R}_+$ & $\vev{\s_R}_-$ & $\vev{|\s_R|}$ \\ \hline
4 & 3.505(52) & -3.518(52) & 3.512(14) \\
8 & 3.337(49) & -3.353(48) & 3.3456(87) \\
16 & 3.342(60) & -3.349(41) & 3.3471(45) \\
\hline
\end{tabular}
\end{center}
\caption{Comparison of subensemble order parameters
to that of the full ensemble, for $(m,g)=(0.10,0.03)$.
\label{sevev}}
\end{table}

\begin{figure}
\begin{center}
\includegraphics[width=2in,height=3in,angle=-90]{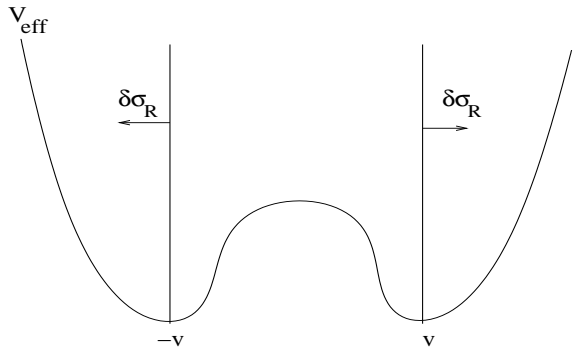}
\end{center}
\caption{Positive fluctuations $\delta \s_{R,m}$ away
from vacua $\pm v$.
\label{veff}}
\end{figure}

This motivates the following fluctuation analysis.
In either $\s_\pm$ domain (connected
regions where $\s >0$ or $\s < 0$) we define $\delta \s_{R,m}$ by
$|\s_{R,m}| \equiv v + \delta \s_{R,m}$.
Thus the fluctuation $\delta \s_{R,m}$ is related
to the value of $\s_{R,m}$ through
\beq
\s_{R,m} &=& v + \delta \s_{R,m}, \quad \s_{R,m} > 0 \nnn
\s_{R,m} &=& -v - \delta \s_{R,m}, \quad \s_{R,m} < 0
\label{ddef}
\eeq
In either case,
the fluctuation is positive if it moves in
the direction that is away from
the origin $\s_{R,m}=0$, relative to $\pm v$; i.e., in the direction
that would increase $|\s_{R,m}|$.  This is shown
in Fig.~\ref{veff}.  In the thermodynamic
limit, $|\delta \s_{R,m}| \ll v$ and $\delta \s_{R,m}$ is
just a fluctuation about one of the two vacua,
which is what we want to study.

Associated with the fluctuation $\delta \s_{R,m}$
one has the following (momentum space) Green function and
susceptibility:
\beq
\tilde G_k(\delta \s_R) = \frac{1}{V} 
\vev{|\delta \tilde \s_{R,k}|^2},
\quad \chi(\delta \s_R) = \tilde G_0(\delta \s_R)
\label{gchi}
\eeq
Here, $V=N_1 N_2$ on an $N_1 \times N_2$ lattice
(typically we take $N_1=N_2=N$) and our convention
for the Fourier transform is
\beq
\delta \tilde \s_{R,k} = \sum_m 
\delta \s_{R,m}
\exp 2\pi i \( \frac{ k_1 m_1}{ N_1}
+ \frac{k_2 m_2}{ N_2} \)
\eeq
A rough estimate of the correlation length
is the one used in \cite{Caracciolo:1994ed}.
It involves the ratio of the Green function 
$\tilde G_k(\delta \s_R)$ at $k=(0,0)$ to $k=(1,0)$:
\beq
\xi(\delta \s_R) = \[ \frac{1}{4 \sin^2 (\pi / N_1)} 
\( \frac{\tilde G_{(0,0)}(\delta \s_R)}
{\tilde G_{(1,0)}(\delta \s_R)} - 1 \) \]^{1/2}
\label{cruxi}
\eeq
Near the critical point we expect this
quantity to exhibit the correct scaling 
behavior, at leading order.
Since we are not interested in high precision
spectral information, \myref{cruxi} will
suffice for our purpose (determining the
location of the critical domain).

We also want to consider fluctuations from
the imaginary part of the field $\s$.
Numerically we have found that $\vev{\s_{I,m}}=0$,
for any real $m,g$.  
We therefore define $\delta \s_{I,m} = \s_{I,m}$ and
\beq
\delta \s_m = \delta \s_{R,m} + i \delta \s_{I,m}
\eeq
This leads to the Green function $\tilde G_k(\delta \s)$, 
susceptibility $\chi(\delta \s)$
and correlation length $\xi(\delta \s)$,
obtained through the replacement $\delta \s_R \to
\delta \s$ in \myref{gchi},\myref{cruxi} above.

\begin{table}
\begin{center}
\begin{tabular}{cccccc}
$m$ & $\vev{|\s_R|}$ & $\xi(\delta \s_R)$ 
	& $\chi(\delta \s_R)$ & $\xi(\delta \s)$ 
	& $\chi(\delta \s)$ \\ \hline
0.00 & 2.360(30) & 7.93(48) & 28.8(3.4)&	12.51(21)&	150.7(4.3) \\
0.03 & 2.462(14) & 7.86(22) & 29.5(1.6)&	12.060(93)&	139.8(1.9)\\
0.05 & 2.624(14) & 7.91(23) & 29.8(1.7)&	11.464(96)&	128.1(2.0)\\
0.10 & 3.512(14) & 7.91(28) & 31.2(2.2)&	9.23(13)&	84.8(2.3)\\
0.15 & 5.012(22) & 6.68(68) & 23.7(4.8)&	6.96(35)&	48.5(4.8)\\
0.20 & 6.6473(89) & 4.85(52) & 12.8(2.7)&	4.89(26)&	25.3(2.7)\\
\hline
\end{tabular}
\end{center}
\caption{Observables on the $N=4$ lattice at $g=0.03$.
\label{csa4}}
%\end{table}

\vspace{0.5in}

%\begin{table}
\begin{center}
\begin{tabular}{cccccc}
$m$ & $\vev{|\s_R|}$ & $\xi(\delta \s_R)$ 
	& $\chi(\delta \s_R)$ & $\xi(\delta \s)$ 
	& $\chi(\delta \s)$ \\ \hline
0.00&	1.593(29)&	11.6(1.0)& 54.3(9.1)& 18.30(41)&	306(12)\\
0.03&	1.8162(94)&	11.08(32)& 57.0(3.2)& 16.68(14)&	262.6(3.8)\\
0.05&	2.0852(99)&	11.53(36)& 62.3(3.8)& 15.13(15)&	221.0(4.2)\\
0.07&	2.5037(98)&	11.17(41)& 61.4(4.4)& 13.17(18)&	170.6(4.6)\\
0.10&	3.3456(87)&	9.72(53)& 48.2(5.1)& 10.06(25)&	103.3(5.2)\\
0.15&	5.0026(60)&	7.21(88)& 22.7(5.3)& 7.17(44)&	45.2(5.4)\\
\hline
\end{tabular}
\end{center}
\caption{Observables on the $N=8$ lattice at $g=0.03$.
\label{csa8}}
%\end{table}

\vspace{0.5in}

%\begin{table}
\begin{center}
\begin{tabular}{cccccc}
$m$ & $\vev{|\s_R|}$ & $\xi(\delta \s_R)$ 
	& $\chi(\delta \s_R)$ & $\xi(\delta \s)$ 
	& $\chi(\delta \s)$ \\ \hline
0.00 & 1.2261(62) & 15.67(49) & 98.3(5.9) & 25.37(19) & 573.7(7.4) \\
0.01 & 1.2309(89) & 15.93(70) & 100.9(8.5) & 25.29(27) & 564(11) \\
0.02 & 1.2802(89) & 15.86(71) & 102.2(8.8) & 24.21(27) & 526(11) \\
0.03 & 1.4076(66) & 16.16(53) & 110.9(7.0) & 22.73(20) & 472.5(8.0) \\
0.05 & 1.7818(96) & 15.88(86) & 116(12) & 18.60(37) & 330(13) \\
0.07 & 2.3579(86) & 13.9(1.1) & 95(14) & 14.56(52)	& 208(15) \\
0.10 & 3.3471(45) & 9.9(1.1)	& 51(11) & 9.89(56) & 101(11) \\
\hline
\end{tabular}
\end{center}
\caption{Observables on the $N=16$ lattice at $g=0.03$.
\label{csa16}}
\end{table}

In Tables~\ref{csa4}-\ref{csa16} we display the measured value
for the observables at $g=0.03$, as a function of $m$,
on three different $N \times N$ lattices, $N=4,8,16$.
A value $g \ll 1$ has been chosen because, as the
reader will recall, $g$ is the coupling in lattice
units.  Thus in terms of the coupling $g_{phys}$ in
physical units, $g=g_{phys} a$.  For $g_{phys}=1$,
the choice $g=0.03$ corresponds to $a=0.03$.  Similarly,
$m=m_{phys} a$, from which one obtains a physical
interpretation of the mass.

Several things are to be noted from the data:
\bit
\item 
The measured value of $\vev{|\s_R|}$ is close to the
classical value $|\s_\pm|=m/g$ in the case $g \ll m$, where
perturbation theory applies.  Thus the sample
is dominated by configurations for which
$\s_m$ sits near one of the two classical
minima, as one would expect.
The classical estimate
breaks down as $m \to 0$.  Note, however, that if
we follow the data at $m=g$ for $N=4,8,16$, the
value of $\vev{|\s_R|}$ is decreasing, and for $N=16$
is only 41\% larger than the classical prediction.
This implies that for sufficiently large $N$,
perturbation theory should be trustworthy
in the regime $g/m \lappeq 1$, as one might guess.
\item
The $\delta \s_R$ observables of susceptibility
and correlation length indicate that the critical
domain is in the neigborhood of $m=0$.  There is
a plateau that is fairly flat (within statistical
uncertainties) extending out from $m=0$ to some
value of $m$ that is not too large.
The range of this plateau appears to be shrinking as $N$
is increased.  From this we conclude that the
critical domain shrinks down to an infinitesmal
neighborhood of $m=0$ in the thermodynamic
limit.  Note that this is consistent with
the nonrenormalization theorem of the continuum
theory, expressed by \myref{wfren}.  That is,
if the bare mass $m$ in \myref{wfren} is set
to zero, the renormalized mass $m_r$ is also
zero:  there is no additive renormalization.
As mentioned above, this can also be viewed as a consequence of
the $U(1)_R$ symmetry, which protects against
a mass term in the superpotential.  The fact
that on the lattice the critical domain is in
a shrinking neighborhood of $m=0$ leads us to
expect that the $U(1)_R$ symmetry holds to a
good approximation for the IR modes in the
continuum limit.  Further evidence
for this will be seen below in Section~\ref{sec:rsym}.
\item
There is a hint of a dip in $\chi(\delta \s_R)$
at $m=0$, but the statistical
errors are too large to be certain.  Simulations
at $m=0$ are very costly, since the leapfrog evolutions
in the hybrid Monte Carlo algorithm become
quite unstable, as discussed in Appendix~\ref{sec:sim}.
In any case, the definition \myref{ddef} of $\delta \s_R$ does
not make much sense right at $m=0$,
since (at least classically) 
there is no barrier and the vacuum is unique.
At $m=0$ it would be more appropriate to study 
fluctuations of $\s_{R,m}$ rather than $|\s_{R,m}|$.
However, since our purpose here is only to find
the critical domain, it is enough to see that the
susceptibility $\chi(\delta \s_R)$, which does
make sense at $m \not= 0$, is rising as we approach 
$m \approx 0$.
\item
The $\delta \s$ observables do not show
a plateau, but are maximized at $m=0$.
The fluctuations $\delta \s_I$ 
contribute significantly, and in fact dominate,
as follows from $\chi(\delta \s) =
\chi(\delta \s_R) + \chi(\delta \s_I)$.
The degree to which this is true increases
as we approach $m=0$.  We will find further
evidence below that $m = 0$ is
the critical point.  As a result, the
$\delta \s$ observables seem to be better
indicators of criticality.
\eit
In conclusion, the critical domain is the neighborhood
of $m=0$.  The continuum renormalization \myref{wfren}
appears to hold to a good approximation.  At $m \gg g$,
the theory is noncritical.

\subsection{R-symmetry of the effective potential}
\label{sec:rsym}

Next we address the two principal questions
of this work:  (1) To what
extent is $Z_2(R)$ a symmetry of the effective
potential?  (2) Is $U(1)_R$ symmetry approximately
recovered in the critical domain?

We introduce a complex constant external field $h$
\beq
\Delta V(h) = - \sum_m \( \bar h \s_m - h \sbar_m \)
\label{vsour}
\eeq
to the scalar potential $V$.
This allows us to explore the effective
potential and to study the extent to
which it is symmetric w.r.t.~$\s \to -\s$,
or the phase rotation $\s \to e^{i\theta} \s$.
As usual, we obtain the effective potential
from the Legendre transformation of the
generating function $w(h)=\ln Z(h)$, where
$Z(h)$ is the partition function that is
obtained when \myref{vsour} is added to the lattice
action.  $Z_2(R)$ symmetry of the effective
potential is equivalent to $w(-h)=w(h)$.
Similarly, $U(1)_R$ symmetry of the effective potential
is equivalent to $w(e^{i \theta} h) = w(h)$.
Note also that $\vev{\s}_h = \p w(h) / \p \bar h$,
where $\vev{\s}_h$ is the expectation value
of $\s$ in the background $h$.
It follows that in the case of $Z_2(R)$ symmetry
we have the prediction
\beq
\vev{\s}_{-h} = - \vev{\s}_h
\label{z2pre}
\eeq
In the case of $U(1)_R$ symmetry we have
the much stronger prediction
\beq
\vev{\s}_{e^{i \theta} h} = e^{i \theta} \vev{\s}_h
\eeq
Equivalently, since we take $m>0,g>0$ and will
find below that $\vev{\s}_h > 0$ if $h$ is real and positive,
\beq
\arg \vev{\s}_h = \arg h, \quad
|\vev{\s}_h| = {\rm const.,} \quad {\rm fixed} ~ |h|
\label{u1con}
\eeq
is the prediction of $U(1)_R$ symmetry.  The second
condition just states that
$|\vev{\s}_h|^2 = \vev{\s_R}_h^2 + \vev{\s_I}_h^2$
is independent of $\arg h$.

\subsubsection{$Z_2(R)$ tests}
\label{z2rs}

Here we consider the case where $h$ is real.
It can be seen from Table \ref{srvh} that, up
to statistical errors, the
values of $\vev{\s_R}$ are supportive of
the $Z_2(R)$ symmetry prediction \myref{z2pre}.
It is a bit surprising that this is already true
on such a small lattice ($N=4$).  Of course
$g/m=0.15$ is well within the perturbative
regime, so upon further reflection it is
reasonable that the irrelevant Wilson mass
interaction terms that break the $Z_2(R)$ symmetry
would only have a mild influence on the
effective potential.  Still, it is remarkable
that at finite lattice spacing the
irrelevant operators have negligible effect.

\begin{table}
\begin{center}
\begin{tabular}{ccc}
$|h|$ & $\vev{\s_R}(h<0)$ & $\vev{\s_R}(h>0)$ \\ \hline
0.10 & -8.427(19) & 8.405(17) \\
0.01 & -6.663(61) & 6.7231(45) \\
0.005 & -5.395(56) & 5.428(53) \\
0.003 & -3.776(79) & 3.897(79) \\
0.001 & -1.408(95) & 1.520(98) \\
0.0005 & -0.667(99) & 0.694(25) \\
0.0001 & -0.235(96) & 0.145(96) \\
0.0 & -0.038(97) & --- \\ \hline
\end{tabular}
\end{center}
\caption{$\vev{\s_R}_h$ as a function of the source $h$
for $(m,g,N)=(0.20,0.03,4)$.  Note that $\vev{\s_I}_h=0$
for $h$ real.
\label{srvh}}
\end{table}

\begin{figure}
\begin{center}
\includegraphics[width=3in,height=5in,angle=90]{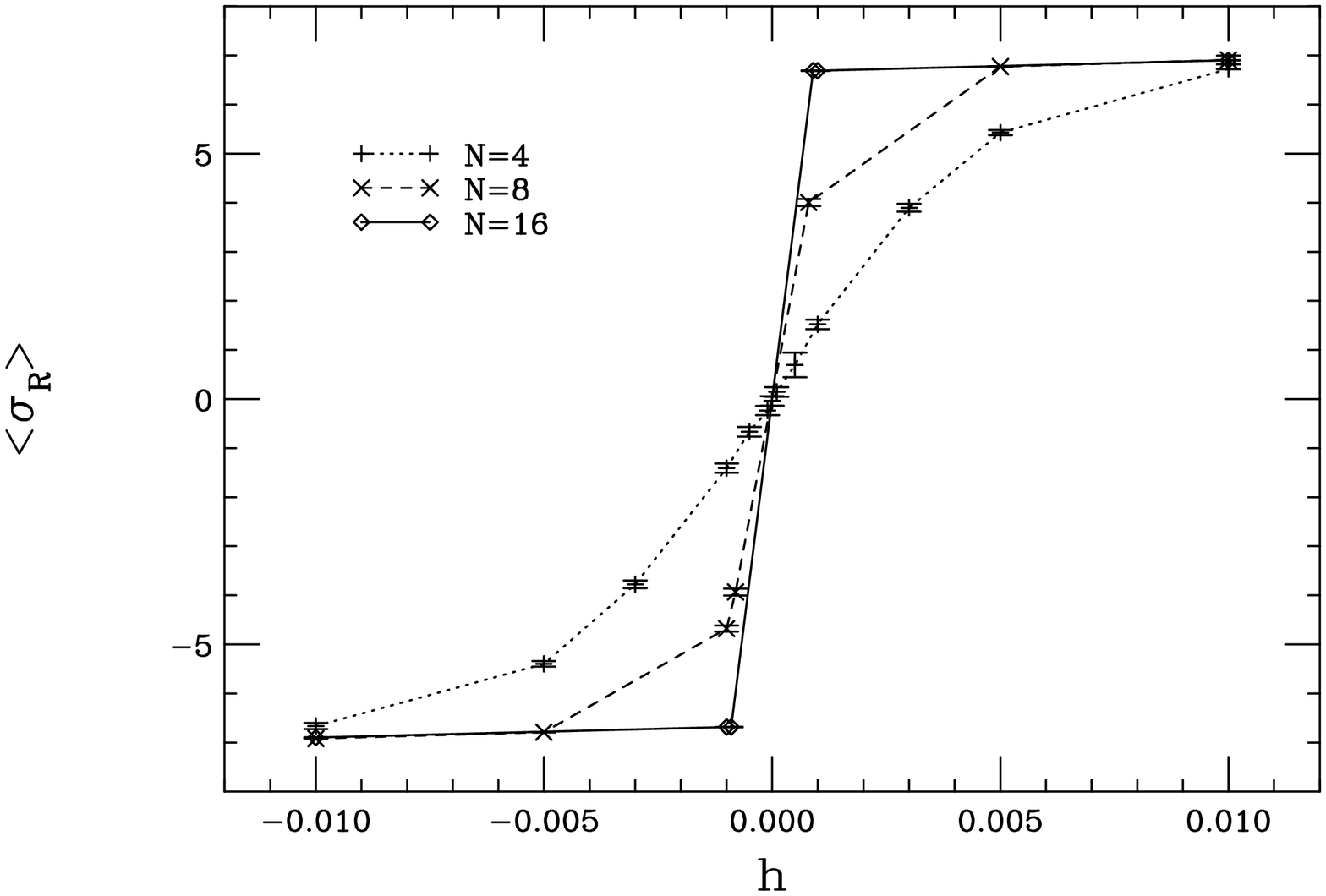}
\end{center}
\caption{$\vev{\s_R}$ versus~$h$ for $(m,g)=(.20,.03)$.
\label{hvm}}
\end{figure}

In Fig.~\ref{hvm} we show the behavior of $\vev{\s_R}_h$
on lattices with $(m,g)=(0.20,0.03)$ and $N=4,8,16$.  
It can be seen that
the transition sharpens significantly as we go
from the $N=4$ lattice to the $N=16$ lattice.
Due to nonergodicity in the simulations
when the barrier between vacua is large, it was not possible to
get stable averages in the transition region,
very close to $h=0$, for the $N=16$ lattice.  
So, we have performed a
hysteresis study at $(m,g,N)=(0.20,0.03,16)$.

\begin{figure}
\centering
\includegraphics[width=3.5in,height=4in,angle=90]{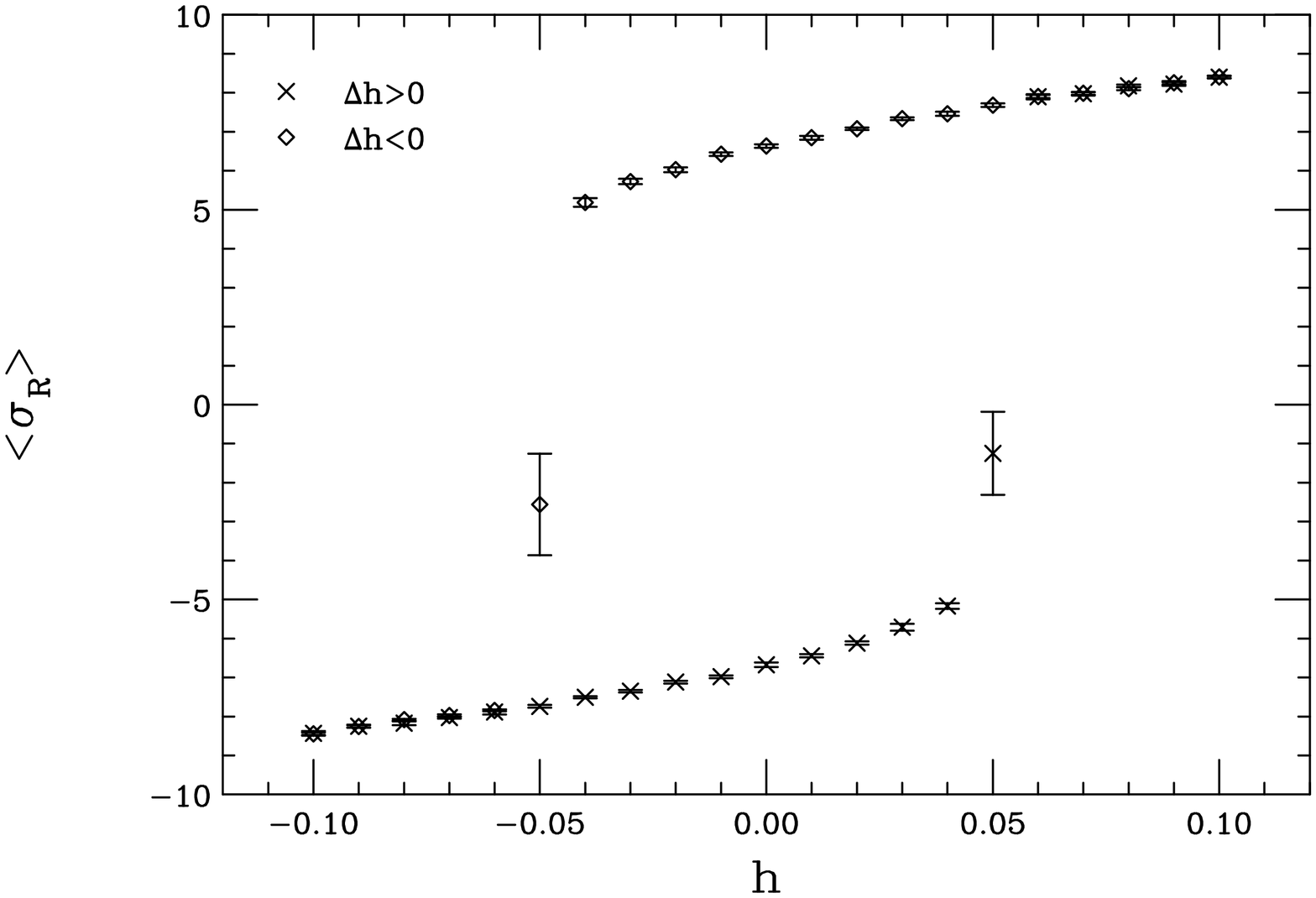}
\\ \vskip 10pt
\includegraphics[width=3.5in,height=4in,angle=90]{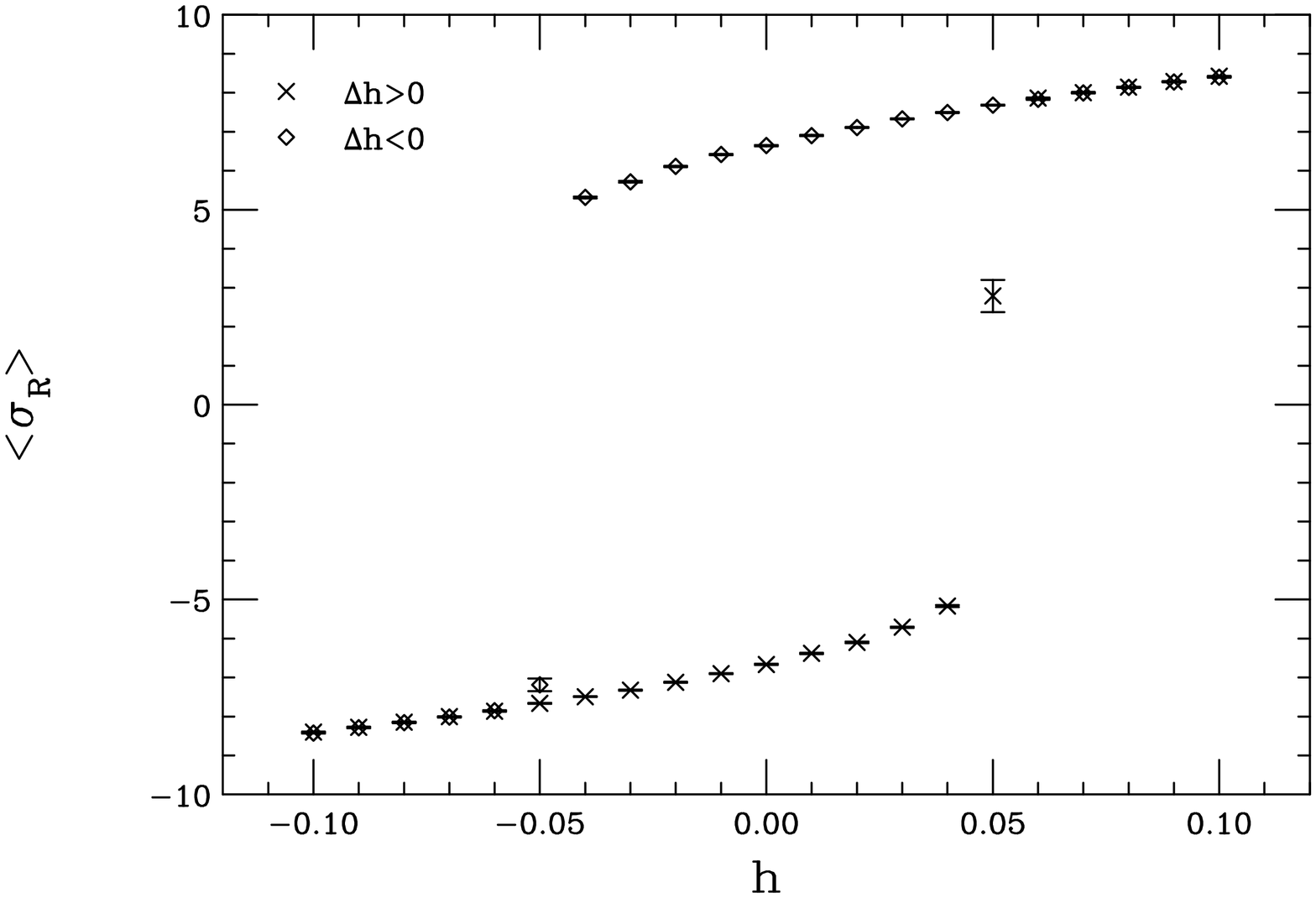}
\caption{Hysteresis cycles for $\vev{\s_R}$ versus~$h$ at
$(m,g,N)=(0.20,0.03,16)$.  The top figure had 20 successive
configurations at each value of 
$h$ along the cycle, whereas the bottom
figure had 200.
\label{hys}}
\end{figure}

A simulation was done where
a thermalized configuration with $h= -0.1$ was
first produced.  Then $\vev{\s_R}$ was computed
from successive configurations obtain
in 20 updates.
Next, $h$ was increased by $\Delta h=0.01$.
Then 20 updates were performed and $\vev{\s_R}$
was calculated from the 20 configurations that
were generated.  This was continued up to $h=0.1$.
Then the cycle was reversed, decreasing $h$
by the same increment.  The results are presented
in Fig.~\ref{hys}.  It can be seen that
there is a pronounced hysteresis.  We increased
the number of successive configurations at each step of the cycle
from 20 to 200 and found that the hysteresis loop
did not tighten at all.  This is indicative that
the transition at $h \approx 0$ is first order.
This is as we should expect from Table~\ref{csa16}:
the point $(m,g)=(0.20,0.03)$ is well outside
the critical domain.  It is only in the critical domain
that the two vacua begin to degenerate, giving
rise to the critical fluctuations which
occur in a second order transition.

Thus, an interesting question is the extent to which
the first order transition softens in the vicinity
of the critical domain $m \approx 0$.  To study this we have produced
hysteresis curves like Fig.~\ref{hys} for a sequence
of decreasing $m$, holding $g=0.03$ fixed.  It can be
seen that the result is entirely consistent with the
conjecture that the critical domain occurs in
the neighborhood of $m=0$.  The hysteresis curves
close up as $m$ is reduced, indicative of the onset
of a second order transition.

Note that the $h=0$ data, $\Delta h > 0$ versus
$\Delta h < 0$, are closer for $m=0$ than for
$m=0.05$.  This supports the conclusion that $m=0$
is the critical point, or the center of the
critical domain.  From this we draw a conclusion about
the plateau and possible dip in $\delta \s_R$
observables that was noted above in relation
to Table \ref{csa16}, versus the $m=0$ maximum in
$\delta \s$ observables, also seen in Table \ref{csa16}.
The fact that less hysteresis is observed
at $m=0$ than at $m=0.05$ indicates that
the former point is ``more critical.''  Since
the $\delta \s$ observables are noticeably
peaked at this point, whereas the $\delta \s_R$
observables are not, we conclude once again
that the former observables are better
indicators of criticality.  
Yet more support for these
conclusions will be obtained below.

\begin{figure}
\centering
\includegraphics[width=2.5in,height=2.5in,angle=90]{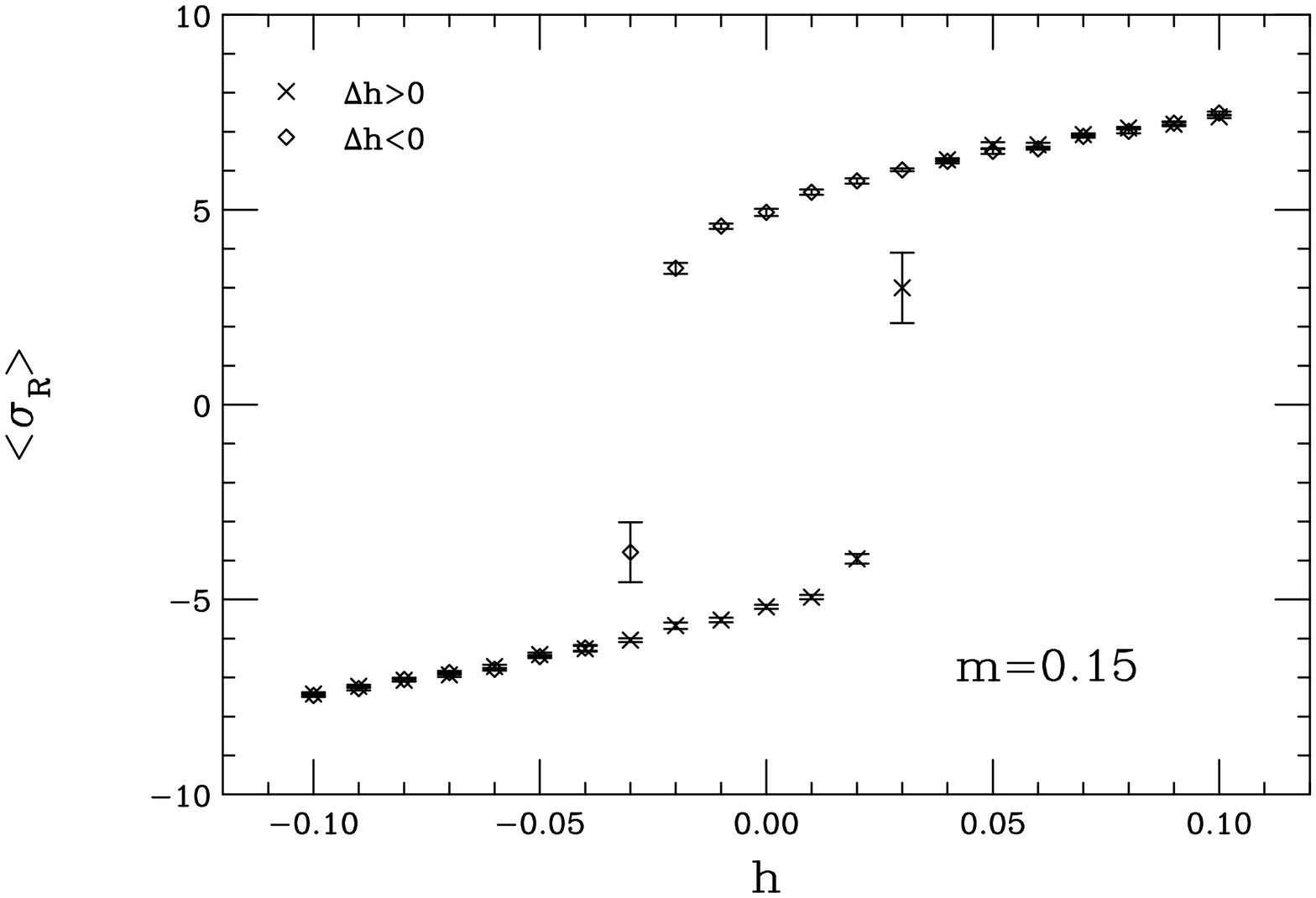}
\hfill
\includegraphics[width=2.5in,height=2.5in,angle=90]{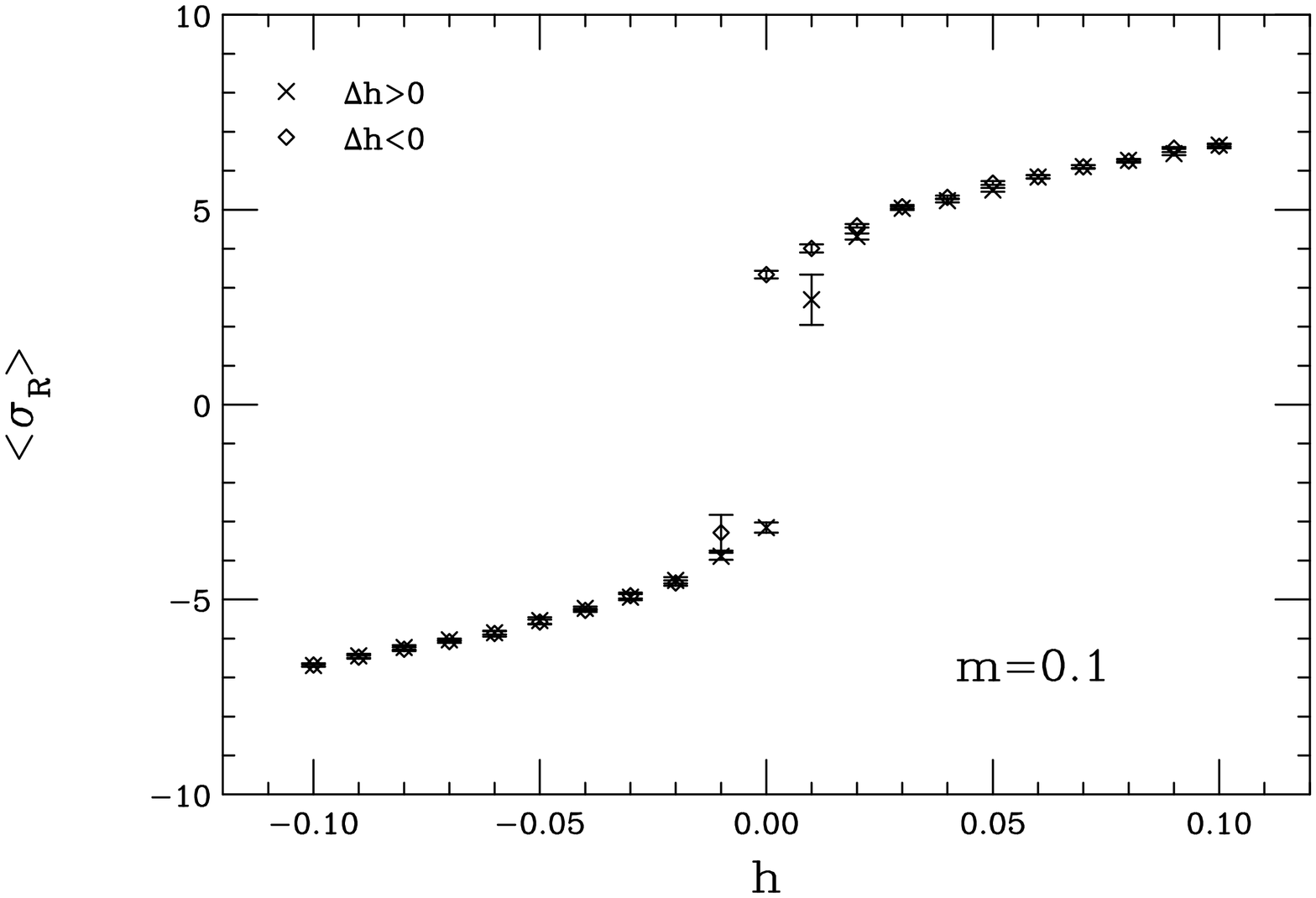}
\\ \vskip 10pt
\includegraphics[width=2.5in,height=2.5in,angle=90]{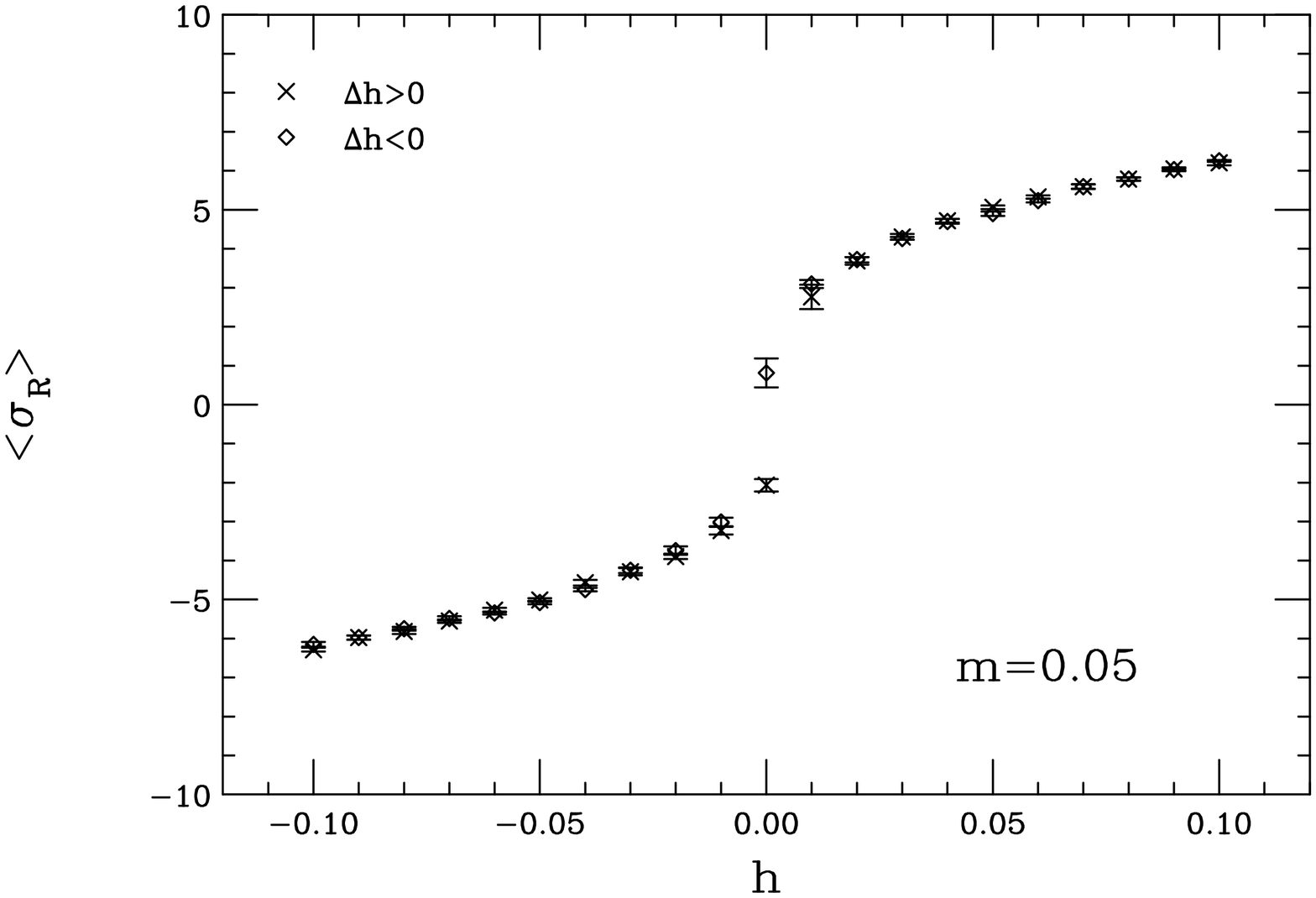}
\hfill
\includegraphics[width=2.5in,height=2.5in,angle=90]{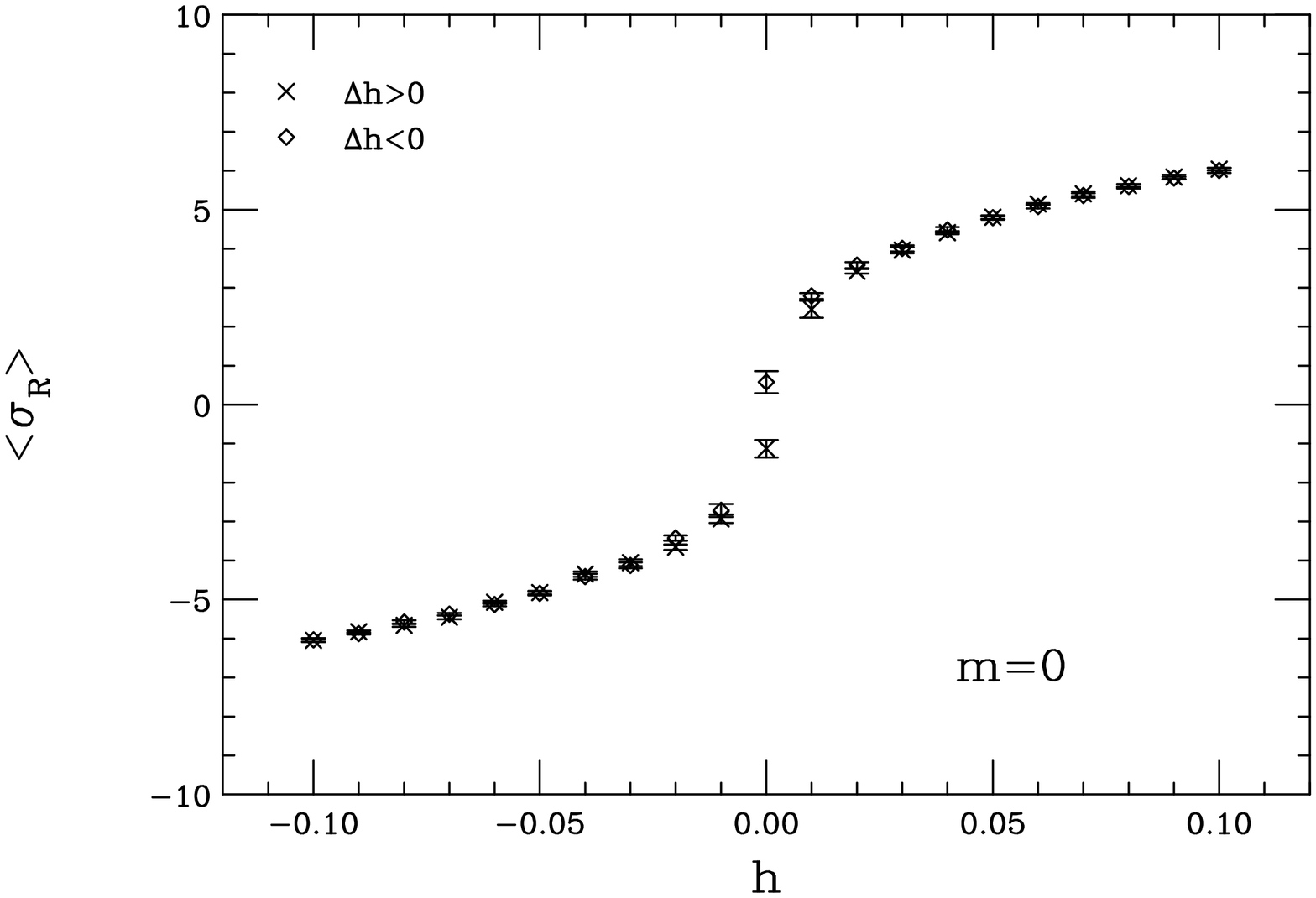}
\caption{Hysteresis cycles for $\vev{\s_R}$ versus $h$ at
$(g,N)=(0.03,16)$, for various values of $m$.  We averaged
over 20 successive configurations 
at each value of $h$ along the cycle.
\label{hysv}}
\end{figure}

\subsubsection{$U(1)_R$ tests}

Our tests of the $U(1)_R$ conjecture \myref{u1con}
were conducted as follows.  Very large ensembles
($10^4$ independent configurations) were generated
at $(g,h,N)=(0.03,0,16)$, for various values
of $m$.  Expectation values $\vev{\s}_h$ at nonzero $h$
were then obtained by reweighting \cite{Falcioni:1982cz}
with the potential \myref{vsour}:
\beq
\vev{\s}_h = \vev{\s e^{- \Delta V(h)}} /
\vev{e^{- \Delta V(h)}}
\label{rwid}
\eeq
The accuracy of the reweighting procedure was
verified by comparison to simulations run at
a nonzero values of $h$.  Note that the
reweighting \myref{rwid} was not used in the
$Z_2(R)$ tests of Section \ref{z2rs} above.
There, simulations included the potential \myref{vsour}.
Comparing to expectation values obtained
there gives us confidence in the reweighting
procedure \myref{rwid}.  

We keep $h$ small for three reasons.  
First, to maintain the reliability of the
reweighting procedure, since a large value of
$h$ will tend to lead to overlap problems;
i.e., the ensemble of configurations simulated
at $h=0$ may be distributed very differently
from one with a large value of $h$.  Second, for
the $N=16$ lattice where we will test for $U(1)_R$,
the volume $N^2$ amplifies the effect of $h$
on the homogeneous mode, $\tilde \s_{k=0} = \sum_m \s_m$,
whose behavior is characterized by
the effective potential.
There is already a large sensitivity 
in $\vev{\s}_h = N^{-2} \vev{\tilde \s_{k=0}}$ 
at very small $h$.  Third, what we are
after are the symmetries
of the effective potential in the $h \to 0$ limit.

\begin{figure}
\begin{center}
\includegraphics[width=4in,height=6in,angle=90]{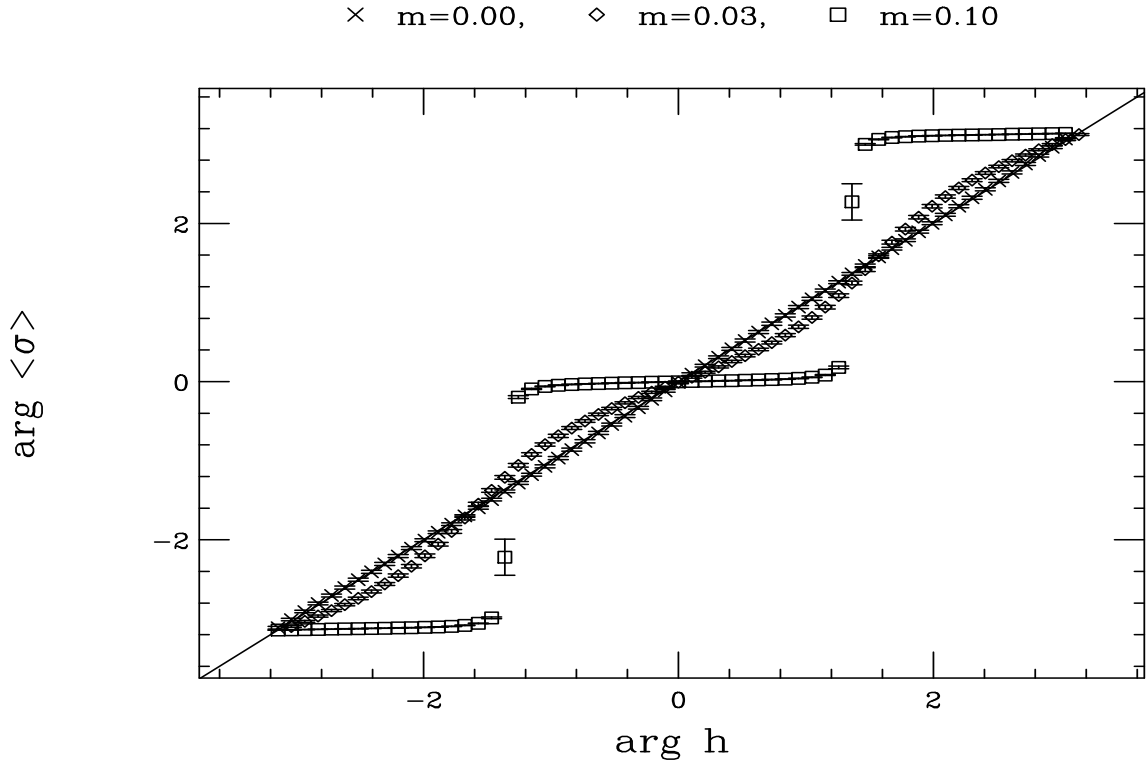}
\end{center}
\caption{A test of $U(1)_R$ symmetry,
by comparison to the prediction $\arg \vev{\s}_h = \arg h$,
indicated by the (diagonal) straight line (mostly
hidden by the $m=0$ data).  It can be seen that
the $m=0$ data is in very good agreement with the
conjecture of $U(1)_R$ symmetry.  
At $m=0.03$, where susceptibilities and
correlation lengths are close to their
critical values, the $U(1)_R$ symmetry
is approximate.  At $m=0.10$,
where from Table \ref{csa16} we see
that we are certainly outside of the
critical domain, the $U(1)_R$ symmetry is
badly broken.  This data is
for $(g,|h|,N)=(0.03,0.001,16)$.
\label{argcomp}}
\end{figure}

In Fig.~\ref{argcomp}
we display $\arg \vev{\s}_h$ versus $\arg h$
at $(g,|h|,N)=(0.03,0.001,16)$ for three
different mass values, $m=0,0.03,0.10$.
For $m=0$, the data passes through the (diagonal) straight
line $\arg \vev{\s} = \arg h$, showing that $U(1)_R$ is a very good
symmetry of the effective potential.  For
$m=0.03$, the data deviates slightly from
the straight line, indicating that
the symmetry is only slightly violated.
Finally, at $m=0.10$, the symmetry is completely
broken.  The fact that $\arg \vev{\s}_h \approx
\pm \pi$ in this case can be understood
as follows.  For larger values of $m$ and
the very small $h$ that we choose, the potential
$V=|W'(\s)|^2$ dominates over the source
potential $\Delta V(h)$ of \myref{vsour}.
In that case, $\vev{\s}_h \approx \pm v$.
Recall that $v$ was defined by \myref{vevhyp},
without the source potential.  Also recall that
$v$ is real for $m,g$ real, as we choose here.  The role
of $h$ then is just as a perturbation
to pick the sign of $\pm v$.
It follows that $\arg \vev{\s}_h \approx \pm \pi$.

The correlation between Fig.~\ref{argcomp}
and the results
of Table~\ref{csa16} is very good:  on the
plateau where $\delta \s_R$ observables
take their critical values, the $U(1)_R$ symmetry
is a good approximation.  At the point
where the $\delta \s$ observables are maximized,
no $U(1)_R$ symmetry violation can be seen
in Fig.~\ref{argcomp} above the statistical
uncertainty.  This indicates that $m=0$
is indeed the critical point, and shows
that $\delta \s$ observables are well-suited
to studying behavior at that point.
We again find that any possible dip
in the $\delta \s_R$ observables at $m=0$
has no significance to the location of
the critical point.

\begin{figure}
\begin{center}
\includegraphics[width=4in,height=6in,angle=90]{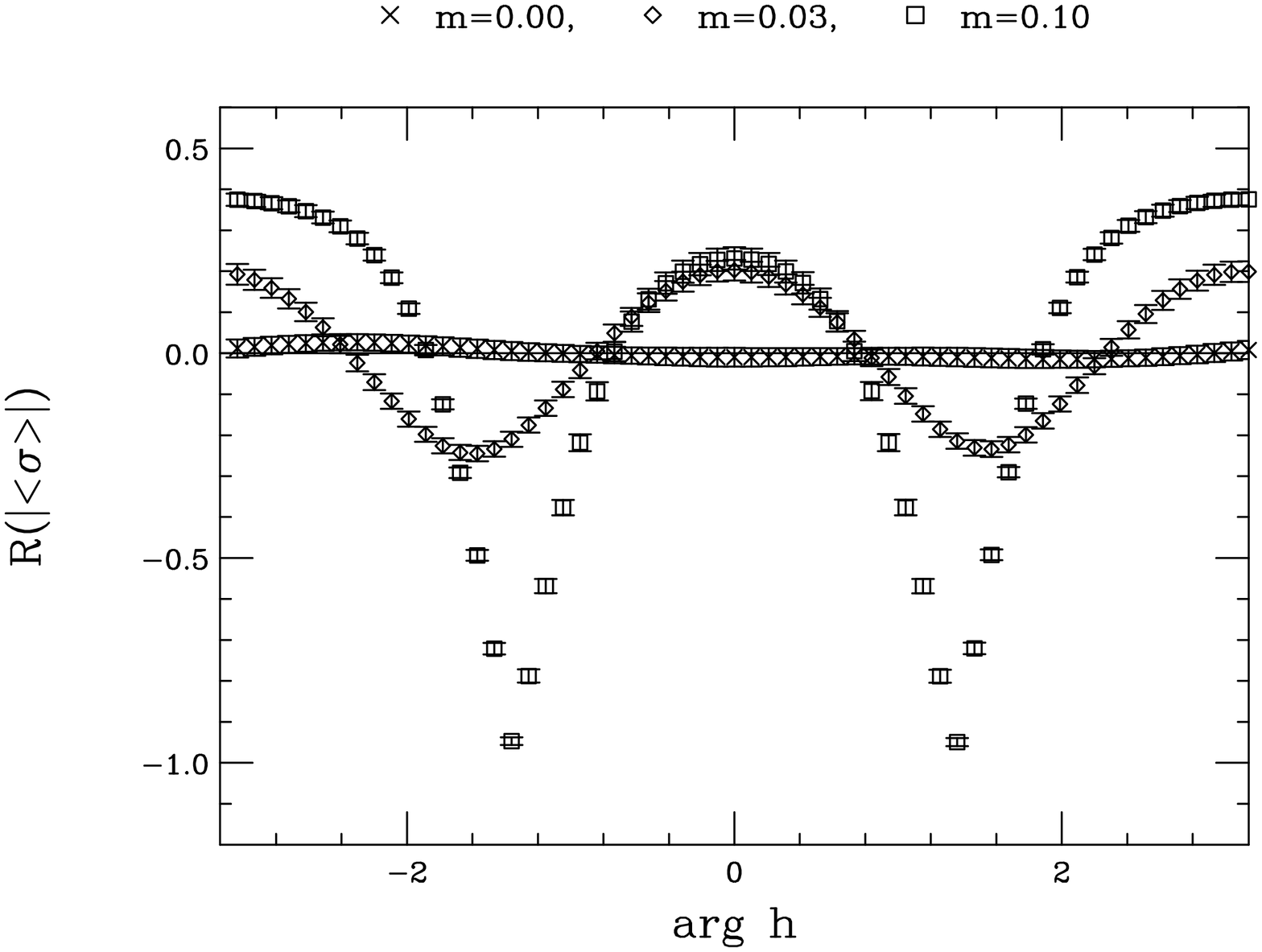}
\end{center}
\caption{A test of $U(1)_R$ symmetry,
by comparison to the prediction that $|\vev{\s}_h|$
should be independent of $\arg h$.  Relative deviation
from the average w.r.t.~$\arg h$ is measured by
$R$ (c.f.~eq.~\myref{Rdf}).
The prediction of $U(1)_R$
symmetry is indicated by the straight line $R=0$.  
Again, the $m=0$ data is in very good agreement with the
conjecture of $U(1)_R$ symmetry, $m=0.03$ data
indicates that the symmetry is approximate (a
maximum of 25\% relative violations),
whereas $m=0.10$ data shows the symmetry
is badly broken.  This data is
for $(g,|h|,N)=(0.03,0.001,16)$.
\label{rmagcomp}}
\end{figure}

The second part of the conjecture \myref{u1con}
was studied through the quantity
\beq
R(|\vev{\s}_h|) = \frac{|\vev{\s}_h|-\overline{|\vev{\s}|}}
{\overline{|\vev{\s}|}}, \quad 
{\overline{|\vev{\s}|}} = \frac{1}{n} \sum_{j=1}^n |\vev{\s}_{h_j}|
\label{Rdf}
\eeq
where $h_j=|h| \exp(2 \pi i j/n)$ corresponds to
the values of $h$ that were used in the data set.
Thus, $R$ measures the relative shift of $|\vev{\s}_h|$ away
from the mean, where the mean is taken w.r.t.~$\arg h$.
In Fig.~\ref{rmagcomp} it can be seen that significant
deviations away from the mean are observed at $m > 0$,
with the largest occuring for the value $m=0.10$.
(The $R$ test of Fig.~\ref{rmagcomp} turns out to be more sensitive
to the $U(1)_R$ violations at $m=0.03$ than
the $\arg \vev{\s}_h$ vs.~$\arg h$ test of
Fig.~\ref{argcomp} was.)  On the
other hand, it can be seen that the $U(1)_R$
symmetry conjecture \myref{u1con} holds up quite
well at $m=0$.  Once again, $m=0$ is found to
be ``more critical'' than $m \approx 0.03$,
and the $\delta \s$ observables seem to
be better indicators of criticality than
the $\delta \s_R$ observables.

In conclusion, the $U(1)_R$ symmetry 
of the effective potential is quite
robust at $m=0$.  To within statistical errors
it is not violated.  This provides strong evidence
that additive renormalization of mass does
not occur in the lattice theory in the continuum
limit.  The violation of the $U(1)_R$ symmetry
in the lattice action at $m=0$ due to the Wilson
mass term in the superpotential does not appear
to impact the renormalization properties of
long distance modes.

\mys{Interpretation and future directions}
\label{s:ifd}

The simulation results that were presented
above are quite encouraging for the Q-exact
action of the (2,2) 2d Wess-Zumino model.
We have found that the R-symmetry of the continuum theory
is a property of the effective potential in the
lattice theory.  The explicit breaking
of the R-symmetry due to the Wilson
mass term in the superpotential is harmless
in the continuum limit; the continuum R-symmetry
is recovered without the need for counterterms.

A very important consequence
of this is that the nonrenormalization theorems
of the continuum theory appear to hold
at a nonperturbative level for the long
distance modes of the lattice theory.
Thus, the good behavior that was already seen
to all orders in perturbation theory \cite{Giedt:2004qs} seems
to persist in the strongly coupled regime 
$g/m \gappeq 1$, where renormalizations
are significant.  Together with the 
supersymmetry Ward identity results of \cite{Catterall:2001wx} and
the spectral results of 
\cite{Beccaria:1998vi,Catterall:2001wx}, the results presented
here provide strong evidence that the nonperturbative
physics of the continuum theory can be
reliably studied with the Q-exact action.
This, in spite of the fact that
it only preserves one out of four
supercharges and explicitly breaks the
R-symmetry, Euclidean invariance, and
reflection positivity that are so 
important to the continuum theory.

Undoubtedly these positive results are related
to (i) the fact that the symmetry breaking is due
to irrelevant operators and (ii) that 1PI diagrams
of UV degree $D \geq 0$ do not occur in 
the lattice perturbation series.
The cancellations of $D=0$ contributions of
subdiagrams in lattice perturbation theory
is intimately related to the exact lattice 
supersymmetry \cite{Giedt:2004qs}.
It would be very interesting to know whether or
not other lattice actions with an exact
supersymmetry, such as the super-Yang-Mills
examples that have been recently proposed
\cite{Cohen:2003xe,Catterall:2004np}, 
have a finite lattice perturbation series,
in the sense that they have no $D \geq 0$
1PI diagrams.  However,
a careful power-counting analysis, comparable
to that done by Reisz for 4d Yang-Mills \cite{Reisz:1988kk}, has yet to
be performed.\footnote{As
is well-known, the UV behavior of the lattice
perturbation series is often worse than
the continuum series, due to modified vertices
and propagators.}
If in these lattice super-Yang-Mills theories 
the perturbation series is finite, the results
presented here give some hope that
those theories have the correct continuum
limit at the nonperturbative level as well.

In research that is in progress,
we are currently subjecting the Q-exact lattice
system studied here to another nonperturbative test
of its continuum limit.  As mentioned
at the outset, the continuum
theory in the critical domain is believed
to afford a Landau-Ginzburg description
of the minimal discrete series of $\Ncal=2$
superconformal field theories.
As a result, the critical exponents of
all relevant operators are known exactly.
If the lattice theory has the correct
continuum limit, it should be able to
reproduce these exponents.  We are
presently studying this issue through
the examination of hyperscaling (dependence
on correlation length) and finite-size 
scaling (dependence on system size)
behavior in the critical regime.  We hope to
report the results of that study in the
near future.  

\newpage

\vspace{20pt}

\noindent
{\bf \Large Acknowledgements}

\vspace{5pt}

\noindent
The author would like to thank Erich Poppitz
for helpful discussions and comments.  Thanks are also
due to Simon Catterall and the Syracuse
Physics Department for use of their Pentium
cluster during some stages of this research.
This work was supported by the National Science and Engineering 
Research Council of Canada and the Ontario 
Premier's Research Excellence Award.

\myappendix

\mys{Simulation details}
\label{sec:sim}

In our simulations, we want to study
the behavior of the theory in the
critical domain.  However, in that case
Monte Carlo simulations tend to exhibit critical
slowing-down due to long wavelength modes that
do not efficiently decorrelate.
The characteristic autocorrelation time 
$\tau_{AC}$ in simulations grows like
\beq
\tau_{AC} \sim |m-m_*|^{-z_m} |g-g_*|^{-z_g}, \quad
z_m,z_g = \ord{1}
\eeq
as we approach a critical point $(m_*,g_*)$.
This is a consequence of the
local trajectories through configuration
space that a typical simulation method
uses.  In some systems, global moves \cite{Swendsen:1987ce} based
on percolation algorithms \cite{perc} can be engineered to
completely overcome this problem.  For example,
in scalar field theories with a $Z_2$ symmetry
one can exploit cluster algorithms that have
been developed for the Ising model 
\cite{Swendsen:1987ce,Montvay:1987us}.
Such a global method
is not known for the present system.\footnote{
It is possible that one might be able
to exploit the approximate $Z_2(R)$ 
symmetry of the effective potential
that is present in this lattice model.  A percolation method
based on this $Z_2(R)$ might work well as a {\it suggestor}
algorithm for Metropolis steps.  I.e.,
a percolation method could be used to
suggest new configurations, which would
be accepted or rejected with a Metropolis
criterion.}  Instead, we have used the method 
of {\it Fourier acceleration} \cite{Davies:1985ad},
as applied \cite{Catterall:2000rv,Catterall:2001jg}
to the {\it hybrid Monte Carlo} (HMC) 
algorithm \cite{Duane:1987de}.
With Fourier acceleration, different Fourier modes are evolved
with disproportionate time steps 
during the {\it leapfrog} trajectories
of the HMC algorithm.  This allows the longest
wavelength modes to be displaced farther
through configuration space, speeding up their
decorrelation.  Simultaneously, shorter wavelength
modes are displaced less far.
This keeps acceptance rates in the Metropolis step
of HMC high.

We have found that Fourier acceleration significantly
reduces autocorrelation times 
for the present system, as previously noted
by Catterall et al.~\cite{Catterall:2001wx}.
At points where we have compared the performance of
conventional leapfrog to Fourier accelerated
leapfrog, we find a reduction of autocorrelation
times by an order of magnitude or more.
Furthermore, with Fourier acceleration applied
to our simulations, we did not encounter problems
with critical slowing-down.  (However, see the
discussion below regarding leapfrog integrator
instabilities in the critical domain.)

In Eq.~(D.16) of~\cite{Giedt:2004qs} it was shown
how to write the fermion matrix in a basis where it is
real.  Once this is done,
we can introduce real pseudofermions
$y_{i,\mbf}, \; i=1,2$ to reproduce the fermion determinant:
\beq
S_P = \half y^T (M^TM)^{-1} y
\eeq
The pseudofermion representation
permits HMC simulation of the system.  We
introduce momenta $p$ and $\pi$, conjugate to $\phi$ and $y$ resp.,
and an auxiliary Hamiltonian:
\beq
H=\bar p p + \half \pi^2 + S_B + S_P
\eeq
With Fourier acceleration, 
the HMC algorithm invokes molecular
dynamics trajectories that evolve
the system according to a modified version
of Hamilton's equations.
The net effect is that the leapfrog
steps are done in Fourier space, with each
mode evolved with its own molecular dynamics
time step.  For further details, see
the discussion in \cite{Catterall:2001jg}.

After the random update of the conjugate momenta
that occurs prior to each 
molecular dynamics trajectory,
we evolve the Fourier modes, labeled by $k$, for
$T$ time steps of spacing $dt_k$.
Thus there is a molecular dynamics evolution
for simulation time $\tau_k = T \cdot dt_k$ between
each randomization of the momenta.
The Fourier acceleration occurs by choosing $dt_k
\propto 1/\omega_k$, where $\omega_k$ is determined
by the frequency of modes in the $g \to 0$ limit,
with some effective {\it accelerator mass} $m_{acc}$,
as described in \cite{Catterall:2001jg}.
Although the random updates of the momenta
introduce noise, in a generalization of the
discrete Langevin equation (see for instance the
very nice discussion in \cite{Kronfeld:1992jf}), it is still
possible that some modes may not decorrelate
(over reasonable periods of simulation time).
This is due to the coincidences pointed
out by Mackenzie \cite{Mackenzie:1989us}.  
To overcome this potential problem, $dt$
was randomized.  We effect this in the
following way.  For a given
number of steps $T$, we choose $dt_{max}$ such
that $T \cdot dt_{max} = \pi/2$.  We then
randomize $dt$ uniformly in $(0,dt_{max})$.
The time step for each mode is then determined
by $dt_k = dt/\omega_k$.  This time step
is used for the entire molecular dynamics trajectory of
$T$ steps.  This algorithm is motivated by the
observation \cite{Kronfeld:1992jf} that in the $g=0$ theory it can
be shown that the autocorrelation of a mode $k$
is proportional to $\cos_T(T dt_k \omega_k)$,
where $\cos_T$ is an approximation to cosine
that becomes exact as $T \to \infty$.  Thus
by choosing $T dt_k \omega_k = \pi/2$ we
get almost complete decorrelation in the $g/m \ll 1$
regime.  In our simulations we find 
rapid decorrelation outside of this regime as well.

We have measured autocorrelation times
throughout all of our simulations.  We
find that
\beq
\tau_{AC} = \ord{1} \times T
\eeq
That is, a few hybrid molecular dynamics trajectories
are sufficient to decorrelate the configurations.
(Hybrid refers to the randomization of momenta
prior to each trajectory.)
Non-ergodicitites were excluded by verifying
that measured quantities were independent of
initial conditions, and that subensembles gave
identical results, within statistical uncertainties.

Unfortunately, as we approach the critical
point $m=0$, the number of steps $T$ in
the leapfrog evolution must typically be chosen 
very large.  Recall that $T dt_{max}$ is held
fixed.  If $dt_{max}$ is too large, we find that
the leapfrog integrator can become 
unstable \cite{Joo:2000dh,Kennedy:2004ae}.
Thus in practice we choose $dt_{max}$ as large
as leapfrog stability will allow, and
adjust $T$ such that $T dt_{max} = \pi / 2$.  In the
critical domain, acceptance rates are then found to
be nearly 100 percent.  (Therefore the only purpose
of the Metropolis step of HMC is to keep the algorithm
``perfect.'')  In practice we find that
at or near $m=0$, it is necessary
to have $T=\ord{10^3}$.  Thus if simulation
time is measured in units of $dt_{max}$,
we do not escape critical slowing down; this
is because $dt_{max}$ must be drastically reduced
as the degree of criticality (measured, say, by
the correlation length in units of lattice spacing)
is increased.
A significant speed-up in the simulations could presumably
be achieved through the multi-pseudofermion methods discussed
in \cite{Kennedy:2004ae}.

\end{document}